\documentclass[letterpaper]{JHEP3}
\pdfoutput=1

\usepackage{amsmath}
\usepackage{amssymb}
\usepackage{amsfonts}
\usepackage{lscape, graphicx}
\usepackage{cite}       
\usepackage{bm}         
\usepackage{verbatim}

\usepackage[all]{xy}
\advance\parskip 1.0pt plus 1.0pt minus 2.0pt   
\def\href#1#2{#2}	

\def\R{{\mathbb R}}
\def\S{{\mathbb S}}

\def\tr{{\rm tr}}

\def\Z{{\mathbb Z}}

\def\Dslash{{\rlap{\raise 1pt \hbox{$\>/$}}D}}

\newcommand{\beq}{\begin{equation}}
\newcommand{\eeq}{\end{equation}}
\newcommand{\beqa}{\begin{eqnarray}}
\newcommand{\eeqa}{\end{eqnarray}}

\def\ltap{\ \raise.3ex\hbox{$<$\kern-.75em\lower1ex\hbox{$\sim$}}\ }
\def\gtap{\ \raise.3ex\hbox{$>$\kern-.75em\lower1ex\hbox{$\sim$}}\ }
\def\gl{\ \raise.5ex\hbox{$>$}\kern-.8em\lower.5ex\hbox{$<$}\ }
\def\roughly#1{\raise.3ex\hbox{$#1$\kern-.75em\lower1ex\hbox{$\sim$}}}

\title{The abelian confinement mechanism revisited: new aspects of the Georgi-Glashow model}

\author
{
    {
    Mohamed M. Anber \footnote{\email{manber@physics.utoronto.ca}}
           \\{Department of Physics, University of Toronto,
    Toronto, ON M5S 1A7, Canada}
           
            }
    }%
    
    \abstract{
 The confinement problem remains one of the most difficult problems in theoretical physics. An important step toward the solution of this problem is the Polyakov's work on abelian confinement. The Georgi-Glashow model is a natural testing ground for this mechanism which has been surprising us by its richness and wide applicability.  In this work, we shed light on two new aspects of this model in $2+1$ D. First, we develop a many-body description of the effective degrees of freedom. Namely, we consider a non-relativistic gas of W-bosons in the background of monopole-instanton plasma. Many-body treatment is a standard toolkit in condensed matter physics. However,  we add a new twist by supplying the monopole-instantons as external background field.  Using this construction, we  calculate the exact form of the potential between two electric probes as a function of their separation. This potential is expressed in terms of the Meijer-G function which interpolates between logarithmic  and linear behavior at small and large distances, respectively. Second, we develop a systematic approach to integrate out the effect of the W-bosons at finite temperature in the range $0 \leq T <M_W$, where $M_W$ is the W-boson mass, starting from the full relativistic partition function of the Georgi-Glashow model. Using a heat kernel expansion that takes into account the non-trivial thermal holonomy,  we show that the  partition function describes a three-dimensional two-component Coulomb gas. We repeat our analysis using the many-body description which yields the same result and provides a check on our formalism. At temperatures close to the deconfinement temperature, the gas becomes essentially two-dimensional recovering the partition function of the dual sine-Gordon model that was considered in a previous work.
    
    \smallskip
    
       {\small{
     }

}}

\begin{document}

\maketitle

\section{Introduction}

After almost 60 years since Yang and Mills formulated their theory \cite{Yang:1954ek}, color confinement remains one of the greatest puzzles in theoretical physics. Eventhough  lattice gauge theories have been successful in demonstrating quark confinement in computer simulations, it is safe to say that up to date there is no analytical understanding of the confinement mechanism in $3+1$ dimensions.
\footnote{See \cite{Greensite:2011zz} for a recent review of the confinement problem, and \cite{Shifman:2008yb} for the elements of the big picture.}
A breakthrough idea toward the solution of the confinement problem was introduced  in the pioneering work of Polyakov \cite{Polyakov:1975rs}, who showed that the proliferation of monopole-instantons in the vacuum of compact QED in $2+1$ dimensions leads to the confinement of electric charges. These monopoles are solution to the Euclidean classical equations of motion, and result due to the compact nature of the $U(1)$ gauge group.  Immediately after this work, Polyakov showed that the same confinement mechanism is at work in the Georgi-Glashow model
\footnote{This model was proposed by Georgi and Galshow in the early seventies of the last century in the context of unified field theories \cite{Georgi:1972cj}.}
 in $2+1$ D \cite{Polyakov:1976fu}. This model consists of an $SU(2)$ Yang-Mills theory coupled to a triplet of scalar fields.
\footnote{ The generalization to $SU(N)$ gauge groups was worked out in \cite{Snyderman:1982ux}.}
Previously, it was shown in  \cite{'tHooft:1974qc,Polyakov:1974ek} that this model admits finite non-singular and nonperturbative solutions to the classical equations of motion. These are the Polyakov 't-Hooft magnetic monopoles in $3+1$ D, and monopole-instantons in the Euclidean setup in $2+1$ D. The monopole-instantons are charged under abelian $U(1)$ group. This is the unbroken compact subgroup of the original $SU(2)$, which breaks spontaneously upon giving a vacuum expectation value to the scalars. The monopole-instantons interact via Coulombic forces and form a gas of monopole plasma. In order to test the effect of the monopole proliferation on two external electric test charges, one computes the behavior of the Wilson loop in the background of the monopole plasma. The Wilson loop $W(C)$ is a gauge invariant order parameter for confinement. Usually, we take our loops $C$ to be rectangles $C=R\times T$ which represent the creation, propagation and annihilation  of two charged probes separated a distance $R$ for time $T$.  In the confinement phase, the expectation value of the Wilson loop experiences an area-law behavior
\begin{eqnarray}
\left\langle W(C) \right\rangle=\left\langle e^{i\oint_C dx_\mu A_\mu } \right\rangle=e^{-\sigma A}\,, 
\end{eqnarray}   
where $A=RT$ is the area of the loop enclosed by the curve $C$, and $\sigma$ is a proportionality constant interpreted as the string tension. In order to understand the area-law behavior, we can think of the Wilson loop as a current loop which itself generates a magnetic field. The monopole and anti-monopoles will line up along the area of the rectangular loop to screen out the generated magnetic field. Therefore, the area-law behavior is associated with the screening sheet of monopoles along the area of the loop. 

A complementary way of thinking about the confinement mechanism is to consider the dual superconductivity picture advocated by 't Hooft and Mandelstam \cite{'tHooft:1995fi,Mandelstam:1974vf}.  In superconductors, the condensation of electric charges, known as Cooper pairs,  breaks the $U(1)$ of electromagnetism spontaneously which in turn gives a mass to the photon. This is the Meissner effect which is responsible for screening the magnetic flux lines in superconductors. However, in type II superconductors magnetic field lines are  allowed to exist in the form of flux tubes known as Abrikosov vortices \cite{Abrikosov:1956sx}. If we place two magnetic monopoles in the bulk of a type II superconductor, the magnetic flux lines can not spread everywhere in the bulk. Instead, they collimate into a thin flux tube which can be though of  a string connecting the two monopoles. Hence, at distance $R$ much larger than the screening length of the superconductor the potential between the probes will behave as $V=\sigma R$, where $\sigma$ is the string tension. According to 't Hooft and Mandelstam, the vacuum of Yang-Mills behaves as type II superconductor except with a reversal of the  rules of electric and magnetic charges. In fact, the Polyakov works \cite{Polyakov:1975rs,Polyakov:1976fu} are the first demonstration of the dual superconductivity picture. Later on, a beautiful implementation of this picture was worked out by Seiberg and Witten in $3+1$ D in a supersymmetric context \cite{Seiberg:1994rs,Seiberg:1994aj}.    

Since the pioneering work of Polyakov, the Georgi-Glashow model has been a testing ground not only for the confinement phenomena, but also for the deconfinement transition. Pure Yang-Mills theory in $3+1$ D experiences a deconfinement transition at strong coupling which hinders a full understanding of the transition \cite{Gross:1980br}. Hence, one needs a simpler theory that resembles the original one, yet under analytic control.  The finite temperature effects in the $2+1$ D Georgi-Glashow  model were first considered in \cite{Agasian:1997wv}.  There, it was shown that a confinement-deconfinement transition happens at temperature $T_c=g_3^2/(2\pi)$, where $g_3$ is the Yang-Mills coupling constant. However, the authors in \cite{Agasian:1997wv} ignored the effect of the W-bosons which plays an important role near the transition region, as was shown later on in \cite{Dunne:2000vp}. Taking the W-bosons into consideration, the authors in \cite{Dunne:2000vp} argued that the partition function near the transition region is that of a two-dimensional double Coulomb gas of monopole-instantons and W-bosons. This gas can be mapped into a dual sine-Gordon model which is further studied using bosonization/fermionization techniques. With such technology, it was shown that the inclusion of the W-bosons modifies the transition temperature to $T_c=g_3^2/(4\pi)$ and that the transition is second order and belongs to the 2D Ising universality class.
\footnote{Also see \cite{Kogan:2002au,Antonov:2004kj} for reviews.}

In this paper, we shed light on some issues that have not been considered before in the Georgi-Glashow model. In the first part of this work, we answer an important question which concerns the  behavior of the potential between two external electric probes at intermediate distances. At distances much shorter then the screening length of the monopole-instanton plasma ${\cal M}^{-1}$, the potential between the probes is logarithmic. On the other hand, at distances much larger than ${\cal M}^{-1}$, the potential is linear. However, an analytic expression for the behavior of the potential in the intermediate region  between the logarithmic and linear potential is still lacking. For this purpose, we develop a Euclidean many-body description of the partition function of the system. In this formalism, we consider the external electric probes as well as W-bosons in the background of the field generated by an arbitrary number of monopole-instantons. Since we are interested only in temperatures much lower than the mass of the W-bosons, we can limit ourselves to a non-relativistic description. Many-body treatment is a powerful tool in condensed matter systems. In the present work, we adapt this method to take into account the effect of monopole-instantons which, to the best of our knowledge, has not been considered elsewhere. At zero temperature, the W-bosons do not play any role thanks to the Boltzmann suppression factor $e^{-M_W/T}$. In this case, we can work only with a partition function that describes the external probes in the background of monopole-instanton gas. Further, we assume that the density of the instantons obey a Gaussian distribution. Although this introduces an error in the string tension, such an assumption considerably simplifies  our computations.  We find that the potential between two probes  of charges $\pm Q$ separated a distance $R$ is expressed in terms of the Meijer-G function 
\begin{eqnarray}
V(R)=\frac{Q^2}{2\pi}\left(\log R+\frac{1}{4}G^{2,3}_{3,5}\left[\left.\begin{array}{ccccc} 1,&1,&\frac{3}{2}& & \\1, & 1,&0,&0,&\frac{1}{2} \end{array}\right|\frac{{\cal M}^2R^2}{4} \right] \right)\,,
\label{The potential in terms of Meijer G function in the introduction}
\end{eqnarray}
which smoothly transits from logarithmic behavior at distances $R<<{\cal M}^{-1}$ to a linear potential $V(R) \rightarrow Q^2 {\cal M}R/4$  at distances much larger than the screening length. This behavior is illustrated in Figure (\ref{The potential and electric field profile between two external probes}) where we  plot the potential and electric field as a function of the separation distance $R$.
 \begin{FIGURE}[ht]
    {
    \parbox[c]{\textwidth}
        {
        \begin{center}
        \includegraphics[angle=0, scale=.8]{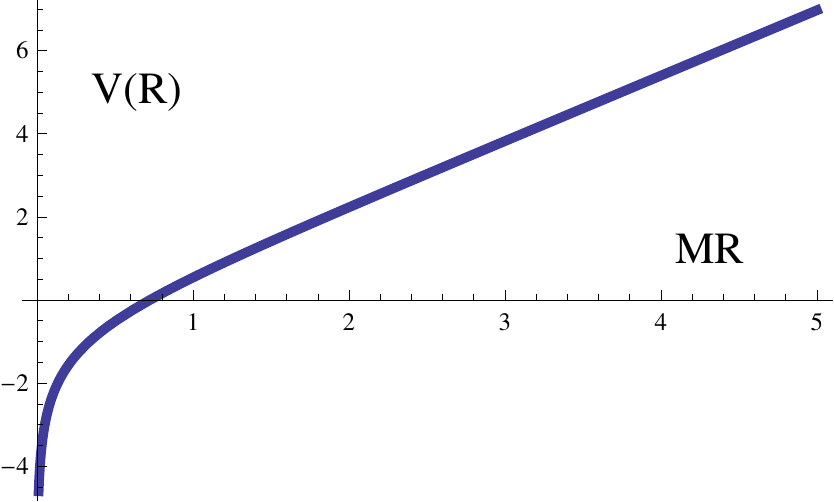}
        \includegraphics[angle=0, scale=.8]{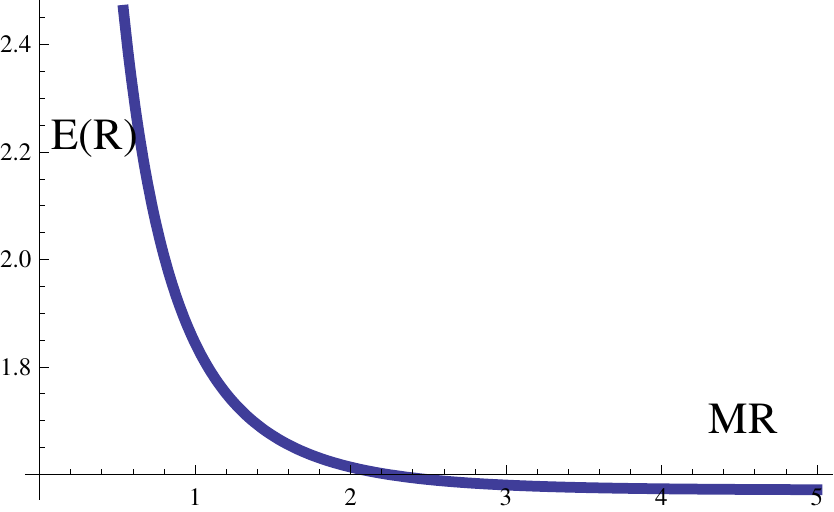}
	\hfil
        \caption 
      {The potential $V(R)$ and electric field $ E=dV(R)/dR$ profiles using the expression (\ref{The potential in terms of Meijer G function in the introduction}). The potential is logarithmic for  distances $R<<{\cal M}^{-1}$, and linear for distances $R>>{\cal M}^{-1}$. The transition from logarithmic to linear behavior happens at a distance $R \sim 3{\cal M}^{-1}$ where  the electric field profile levels off.
			}
      \label{The potential and electric field profile between two external probes}
       \end{center}
        }
    }
\end{FIGURE}

In the second part of our work, we perform a systematic study of the finite-temperature partition function of the Georgi-Glashow model in the temperature range $0 \leq T < M_W$, where $M_W$ is the W-boson mass. In this regard, we take two different approaches.  In the first approach, we start our treatment  from the full relativistic partition function which treats the monopole-instantons as external background field, and then apply a heat kernel expansion technique which takes into account the non-trivial thermal holonomy. This results in an effective action that contains both relevant and irrelevant operators. Ignoring the irrelevant ones, we show that the partition function of the system takes the form of the grand canonical distribution  of a non-relativistic three-dimensional double Coulomb gas:
\begin{eqnarray}
\nonumber
{\cal Z}_{\mbox{\scriptsize grand}}&=&\sum_{N_{m\pm}, q_a=\pm 1 }\sum_{N_{W\pm}, q_A=\pm 1 }\frac{\xi_m^{N_{m+} + N_{m-}}}{N_{m+}! N_{m-}!}\frac{(T\xi_W)^{N_{W+} + N_{W-}}}{N_{W+}! N_{W-}!}\prod_a^{N_{m+} + N_{m-}} \int d^{2+1}x_a\prod_A^{N_{W+} + N_{W-}} \int \frac{d^{2}x_A}{T}\\
\nonumber
&&\times \exp \left[-\frac{8\pi^2}{g_{3}^2}\sum_{a,b}q_aq_bG(x_a- x_b)+\frac{g_{3}^2}{4\pi T}\sum_{A,B}q_Aq_B\log T|\vec x_A-\vec x_B|+2i\sum_{aA}q_aq_A\Theta\left(\vec x_a-\vec x_A\right)\right]\,.\\
\label{final expression for Z in the introduction}
\end{eqnarray}
 The gas consists of W-bosons and monopole-instantons with fugacities $\xi_W$ and $\xi_m$, respectively. Two W-bosons carrying  charges $q_A=\pm 1$ and located at $\vec x_A$  interact logarithmically (which is the two-dimensional potential) at all temperatures in the range $0 \leq T < M_W$. While two monopole-instantons with charges $q_a=\pm 1$ and three-dimensional positions $x_a$ interact via $G(x_a-x_b)$, where $G$ is the Green's function of the Laplacian operator on $\R^2\times \S^1_\beta$, and $\S^1_\beta$ is the thermal circle. In addition, W-bosons interact with monopole-instantons via the  Aharonov-Bohm phase  $q_aq_A\Theta\left(\vec x_a-\vec x_A\right)$.  At temperatures much larger than the inverse distance between two-monopole instantons, yet much lower than the deconfinement temperature, the  Green's function $G$ reduces to a logarithmic function and the gas becomes essentially two-dimensional recovering the same partition function considered before in \cite{Dunne:2000vp}. In the second approach, we start from the non-relativistic many-body description and then integrate out the W-bosons to recover (\ref{final expression for Z in the introduction}). This also works as an independent check which ensures the validity of our many-body description.
 
This paper is organized as follows. In Section 2, after a quick review of the Georgi-Glashow model, we write down the relativistic partition function of the system taking into account the monopole-instantons as background field. Then, motivated by a physical picture, we give a non-relativistic many-body description of the system. In Section 3, we use the many-body partition function, aided by a mean-field approach, to derive the potential between two static electric probes at zero temperature. In Section 4, we use both the relativistic and  non-relativistic partition functions to show that the finite-temperature Georgi-Glashow model can be thought of as a three-dimensional double Coulomb gas as explained above. Finally, we conclude in Section 4 and provide directions for future research. The paper contains three appendices displaying miscellaneous calculations.

\section{Theory and formulation}

\subsection{The Georgi-Glashow model: perturbative treatment}

We consider the Lagrangian of  $SU(2)$ Georgi-Glashow model in $\R^{2,1}$ 
\begin{eqnarray}
{\cal L}=-\frac{1}{4g_3^2}F_{\mu\nu}^aF^{a\,\mu\nu}+\frac{1}{2}\left(D_\mu\phi^a\right)\left(D^\mu\phi^a\right)-\lambda\left(\phi^a\phi^a-v^2\right)^2,
\label{georgi-glashow}
\end{eqnarray}
where $g_3$ is the three-dimensional coupling constant, and $\phi^a$ are matter fields in the adjoint representation of the $SU(2)$ group. The Greek indices $\mu$ run from $0$ to $2$ and the color indices $a$ run from $1$ to $3$. The field strength tensor $F^a_{\mu\nu}$ and the covariant derivative $D_\mu$ are given by
\begin{eqnarray}
\nonumber
F_{\mu\nu}^a&=&\partial_\mu A^a_\nu-\partial_\nu A^a_\mu+\epsilon^{abc}A_\mu^bA_\nu^c\,,\\
D_\mu \phi^a&=& \partial_\mu\phi^a+\epsilon^{abc}A_\mu^b \phi^c\,.
\end{eqnarray}
The field $\phi^a$ acquires a vacuum expectation value $v$ (say in the $a=3$ direction) which breaks the $SU(2)$ down to $U(1)$. Then, we write the third component $\phi^3$ of the Higgs field as $\phi^3=v+\phi$, where $\phi$ is the physical excitation of the Higgs field. The third color component of the gauge field  $A^{3}_\mu\equiv A_\mu$ remains massless, and the other two components form massive vector bosons:
\begin{eqnarray}
W_\mu^{\pm}=\frac{1}{\sqrt{2}g_3}\left(A_\mu^1 \pm i A_\mu^2\right)\,.
\label{charged W}
\end{eqnarray} 
The massive vector bosons, or W-bosons for short, carry electric charge $\pm e$, where $e \equiv g_3$. We also define the charged Goldstone Bosons $\phi^{\pm}$ as
\begin{eqnarray}
\phi^\pm=\frac{1}{\sqrt{2}}\left(\phi^1 \pm i \phi^2\right)\,.
\label{Goldston bososns}
\end{eqnarray}
Substituting (\ref{charged W}) and  (\ref{Goldston bososns}) into (\ref{georgi-glashow}), we find that the Lagrangian (\ref{georgi-glashow}) can be rewritten as the sum of quadratic and interaction pieces ${\cal L}_{\mbox{\scriptsize quad}}+{\cal L}_{I}$:
\begin{eqnarray}
\nonumber
{\cal L}_{\mbox{\scriptsize quad}}&=&-\frac{1}{4g_3^2}F_{\mu\nu}F^{\mu\nu}-2i F^{\mu\nu}W_\mu^+W_\nu^--\left(D^{+\mu}W^{+\nu}\right)\left(D_\mu^-W_\nu^-\right)+g_3^2v^2W_\mu^-W^{+\mu}\\
\nonumber
&&+\frac{1}{2}\partial_\mu \phi \partial^\mu \phi+\frac{1}{2}m_H^2\phi^2+\left(D_\mu^+ \phi^+ \right)\left(D^{-\mu} \phi^- \right)\\
&&+\left(D^{-\mu}W_\mu^-\right)\left(D^{+\nu}W_\nu^+\right)+ig_3vW_\mu^- D^{+\,\mu}\phi^+-ig_3vW_\mu^+ D^{-\,\mu}\phi^-\,,
\label{quad lagrangian }
\end{eqnarray}
and
\begin{eqnarray}
\nonumber
{\cal L}_{\mbox{\scriptsize I}}&=&\frac{g_3^2}{4}\left(W_\mu^+W_\nu^--W_\mu^-W_\nu^+\right)^2+g_3^2(\phi^2+2v\phi)W_\mu^-W^{+\mu}+ig_3\phi W_\mu^- D^{+\,\mu}\phi^+-ig_3\phi W_\mu^+ D^{-\,\mu}\phi^-\\
&&+ig_3W^{+\mu}\phi^-\partial_\mu\phi-ig_3W^{-\mu}\phi^+\partial_\mu\phi-\frac{g_3^2}{2}\left(W^{-\mu}\phi^+-W^{+\mu}\phi^-\right)^2\,,
\end{eqnarray}
where  $F_{\mu\nu}=\partial_\mu A_\nu-\partial_\nu A_\mu$, and $D_\mu^{\pm}=\partial_\mu \pm i A_\mu$. At this stage, let us emphasize that  the field $A_\mu$  does not only describe  the photon fluctuations, but it can also include any external $U(1)$ background fields, like the monopole-instanton field, as we will explain shortly. 

The last line in the quadratic Lagrangian (\ref{quad lagrangian }) has three terms that may add difficulties to our analysis. We can use integration by parts in the first term to get $-W^-_\mu D^{+\mu}D^{+\nu}W_\nu^+$ which is not in the form of a Klein-Gordon operator, i.e. $D_\mu^2$. The other two terms couple the W-bosons to the Goldston bosons. Fortunately, one can get rid of these three terms by entertaining the fact that our Lagrangian has a gauge freedom. Therefore, by choosing an appropriate gauge we can eliminate the unwanted terms. To this end, we add the gauge fixing Lagrangian  ${\cal L}_{\mbox{\scriptsize GF}}=-G^+G^-$, where
\begin{eqnarray}
G^{\pm}=D_\mu^\pm W^{\pm \mu}\mp i g_3v\phi^{\pm}\,.
\end{eqnarray}
Thus
\begin{eqnarray}
{\cal L}_{\mbox{\scriptsize GF}}=-\left(D^{-\mu}W_\mu^-\right)\left(D^{+\nu}W_\nu^+\right)-ig_3vW_\mu^- D^{+\,\mu}\phi^++ig_3vW_\mu^+ D^{-\,\mu}\phi^--g_3^2v^2\phi^+\phi^-\,.
\end{eqnarray}
It is clear that this gauge fixing Lagrangian eliminates the last three terms in (\ref{quad lagrangian }). After gauge fixing, we also need to include the ghost contribution ${\cal L}_{\mbox{ \scriptsize Ghost}}=-c^+\frac{\delta G^-}{\delta\alpha^-} c^-$, where $c^\pm$ are the ghost fields. To calculate the quantity $\frac{\delta G^+}{\delta\alpha^+}$ we proceed as follows. First, we note that the fields $A_\mu^a$ and $\phi^a$ transform under the infinitesimal gauge transformation $\alpha^a$ as $\delta A_\mu ^a=D_\mu \alpha^a$ and $\delta \phi^a=-\epsilon^{abc} \alpha^b \phi^c$. Then, we decompose $\alpha^a$ into $\alpha^3$, the gauge parameter of the unbroken $U(1)$ group, and $\alpha^{\pm}=(\alpha^1\pm i\alpha^2)/\sqrt{2}$ along the $1$ and $2$ color directions. Hence, we obtain
\begin{eqnarray}
\nonumber
\delta \phi^\pm&=&\pm i \alpha^\pm v\mp i\alpha^3 \phi^\pm\,,\\
g_3\delta W_\mu^\pm&=&D_\mu^\pm \alpha^\pm \mp i \alpha^3 g_3 W_\mu^\pm\,.
\end{eqnarray}
Then, we find $\frac{\delta G^-}{\delta\alpha^-}=\frac{1}{g_3}\left(D_\mu^-D^{-\mu}+g_3^2v^2\right)$. 

In the following, it is more appropriate to work in Euclidean space. A Euclidean version of the Lagrangian can be obtained by performing a Wick rotation. Adding the contribution from the gauge fixing and ghost terms, the total Lagrangian reads (from here on, we do not distinguish between upper and lower indices)
\begin{eqnarray}
\nonumber
{\cal L}_{\mbox{\scriptsize total}}&=& {\cal L}_{\mbox{\scriptsize quad}}+{\cal L}_{\mbox{\scriptsize I}}+{\cal L}_{\mbox{\scriptsize GF}}+{\cal L}_{\mbox{\scriptsize Ghost}}  \\
\nonumber
&=&\frac{1}{4g_3^2}F_{\mu\nu}F_{\mu\nu}+\frac{1}{2}\partial_\mu \phi \partial_\mu \phi+\frac{1}{2}m_H^2\phi^2\\
\nonumber
&&+W^+_{\mu}\left(\delta_{\mu\nu}\left( -D_\alpha D_{\alpha}+M_W^2\right)-2iF_{\mu\nu}\right)   W^-_{\nu}+\phi^+\left(- D_\alpha D_{\alpha}+M_W^2\right) \phi^-\\
&&+c^+\left( -D_\alpha D_{\alpha}+M_W^2\right) c^-+\mbox{nonquadratic terms}\,,
\label{total lagrangian}
\end{eqnarray}
where $D_\alpha=\partial_\alpha-iA_\alpha$, and we have used $M_W^2=g_3^2v^2$ where $M_W$ is the W-boson mass. Notice that the Lagrangian ${\cal L}_{\mbox{\scriptsize total}}$ is invariant under the electromagnetic $U(1)$ gauge group since the gauge fixing we have used leaves the $U(1)$ subgroup of the $SU(2)$ intact.  

This ends our treatment of the perturbative part. However, the Georgi-Glashow model contains also nonperturbative solutions. These are monopole-instantos that were first discovered by Polyakov \cite{Polyakov:1974ek} and 't Hooft \cite{'tHooft:1974qc} as solitons in the Georgi-Glashow model in $3+1$ D.  According to the path integral formulation of field theory, the grand partition function of the system is obtained by summing over all trajectories, that take us from one point in the field space to the other, weighted by their action:
\begin{eqnarray}
{\cal Z}=\sum_{\mbox{\scriptsize paths}}e^{-S_{\mbox{\scriptsize path}}}\,.
\label{schematic path integral}
\end{eqnarray}
This sum must include contributions from both perturbative and nonperturbative sectors. Before writing down the partition function of the Georgi-Glashow model, in the following section we review the nonperturbative solutions of the theory at hand.  

\subsection{Nonperturbative effects: adding monopole-instantons}

In addition to the perturbative excitations described above, the three-dimensional Georgi-Glashow model admits nonperturbative objects. These are monopole-instantons allowed by the non-trivial homotopy $\pi_2\left(SU(2)/U(1)\right)=\pi_1\left(U(1)\right)=\Z$, and are obtained as classical solutions to the Euclidean non-abelian equations of motion.  These solutions have to be included in the path integral formulation of the field theory, which can have dramatic effects on the physics.  Monopole-instantons are particle-like objects localized in space and time, have internal structure and mediate long range force, thanks to the unbroken $U(1)$. Although a single instanton solution satisfies the equations of motion, two or more instantons do not. However, if these objects are well separated, then a solution that is a superposition of many instantons can still be a good approximate solution to the equations of motion. In a reliable semi-classical treatment, one  includes an arbitrary number of these objects in the path integral provided that they are well separated, or in other words, their density is low. This is known as the dilute gas approximation. In such approximation, the internal structure of the instantons does not play any role, and for all purposes we can replace the non-abelian field solution with an abelian one.

The abelian field of a single monopole-instanton localized at the origin $\left(x_0=0, \vec x=0\right)$ is given by
\begin{eqnarray}
\nonumber
A_0^m(\vec x,x_0)&=&-\frac{x_1}{r\left(r+x_2\right)}\,,\\
\nonumber
A_1^m(\vec x,x_0)&=& \frac{x_0}{r\left(r+x_2\right)}\,,\\
A_2^m(\vec x,x_0)&=&0\,,
\label{monopole field}
\end{eqnarray}
where $x_{1,2}$ and $x_0$ are respectively the spatial and Euclidean time coordinates, and $r=\sqrt{x_0^2+x_1^2+x_2^2}$ is the spherical-polar radius. The above solution is singular at $x_2=-r$. This is the Dirac string that stems from the location of the monopole-instanton at the origin and extends all the way along the negative $x_2$-axis. This string is not physical; it is just a gauge artifact as can be shown directly by calculating the monopole field $B^m_\mu=\epsilon_{\mu\nu\alpha}\partial_\nu A^m_\alpha=\frac{x_\mu}{r^3}$.  The magnetic charge carried by a single monopole-instanton is defined as the surface integral of the monopole magnetic field over a 2-sphere, divided by $g_3$:
\begin{eqnarray}
Q_m \equiv\frac{1}{g_3}\int_{S^2} dS_\mu B^m_\mu=\frac{4\pi}{g_3}\,.
\end{eqnarray}
In the dilute gas approximation, we add the contribution from an arbitrary number of these monopole-instantons that are randomly distributed all over the spacetime. This results in the total  field
\begin{eqnarray}
{\cal A}_\mu(\vec x, x_0)=\sum_{a}q_aA^m_\mu (\vec x- \vec x_a,x-x_{0a})\,,\quad
{\cal B}_\mu(\vec x, x_0)=\sum_{a}q_aB^m_\mu (\vec x- \vec x_a,x-x_{0a})\,,
\label{total monopole field}
\end{eqnarray}
where $\left(\vec x_a, x_{0a}\right)$ is the position of the monopole-instanton and $q_a=\pm1$ is its charge. Since monopole-instantons carry $U(1)$ charges, they will interact via Coulombic forces. The form of interaction can be obtained from the action
\begin{eqnarray}
S=\frac{1}{4g_3^2}\int d^3 x{\cal F}_{\mu\nu}{\cal F}_{\mu\nu}=4\pi g_m^2\sum_{a>b}\frac{q_aq_b}{|x_a-x_b|}\,,
\end{eqnarray}   
where ${\cal F}_{\mu\nu}=\partial_\mu {\cal A}_\nu-\partial_\nu {\cal A}_\mu$, and we have introduced the magnetic coupling 
\begin{eqnarray}
g_m\equiv \frac{1}{g_3}=\frac{Q_m}{4\pi}\,.
\end{eqnarray}
The electric and magnetic couplings are related by the Dirac quantization condition $eg_m=1$. This is twice the minimal value allowed for $SU(2)$ since the $W$-bosons have twice the minimal charge.

Now, we are ready to include both perturbative and nonperturbative sectors in the path integral sum (\ref{schematic path integral}).

\subsection{The grand partition function}

The grand partition function of the system is obtained as a path integral over the fields $A_\mu$, $W_\mu^\pm$, $\phi^\pm$,  $c^\pm$, and $\phi$. Then, one includes the contribution from the nonperturbative sector as a sum over an arbitrary number of positive $N_{m+}$  and negative $N_{m-}$ monopole-instantons. Hence, the grand partition function reads 
\begin{eqnarray}
\nonumber
{\cal Z}{\mbox{\scriptsize grand}}&=&\sum_{N_{m\pm}, q_a=\pm 1 }\frac{\xi_{m}^{N_{m+} + N_{m-}}}{N_{m+}! N_{m-}!}\left(\prod_a^{N_{m+} + N_{m-}} \int d^3x_a\right) \\
&\times&\int [{\cal D} A_\mu^{\mbox{\scriptsize ph}}][{\cal D}W^-_\mu][{\cal D}W^+_\mu][{\cal D}\phi][{\cal D}\phi^+][{\cal D}\phi^-][{\cal D}c^+][{\cal D}c^-] \exp\left[-\int d^3 x{\cal L}_{\mbox{\scriptsize total}}\right]\,,
\label{total partition function}
\end{eqnarray}
where the monopole fugacity  $\xi_m$ is given by
\footnote{ There is a typo in the pre-exponential factor of $\xi_m$ in the original work \cite{Polyakov:1976fu} which propagated to other places. The correct expression is given in  \cite{Shifman:2012zz}, which we use here. }
\begin{eqnarray}
\xi_m=\mbox{constant} \times M_W^5 g_3^{-4}\exp\left[-4\pi \frac{M_W}{g_3^2}\epsilon\left(\frac{M_W}{m_H}\right)\right]\,,
\label{the monopole fugacity}
\end{eqnarray}
and $\epsilon\left(\frac{M_W}{m_H}\right)$ is a function of the ratio between the W-boson mass and the Higgs mass. This function tends to unity in the Bogomolny-Prasad-Sommerfield (BPS) limit \cite{Prasad:1975kr,Bogomolny:1975de}, $m_H \rightarrow 0$, and tends to $1.79$ in the opposite limit $m_H \rightarrow \infty$ \cite{Kirkman:1981ck}. In the following, we assume that the Higgs mass is heavy and hence the Higgs field is short ranged and we can neglect its effects in our analysis.  
As we mentioned above, in order for the partition function to make sense, the monopole-instanton gas has to be dilute, or in other words $\xi_m<<1$. This in turn requires that we work in the weak coupling limit $g_3/v<<1$. Given the monopole fugacity $\xi_m$, the average distance between two monopole-instantons is $\sim e^{\frac{2\pi v}{3g_3}\epsilon(M_W/m_H)}$, apart from a dimensionfull pre-exponential factor.

It is very important to stress that the field $A_\mu$  appearing in the Lagrangian (\ref{total lagrangian}) is given by $A_\mu=A^{\mbox{\scriptsize ph}}_\mu+{\cal A}_\mu$, and hence $F_{\mu\nu}=F^{\mbox{\scriptsize ph}}_{\mu\nu}+{\cal F}_{\mu\nu}$. Thus, $A_\mu$  includes contributions from both the fluctuations of the dynamical photon $A^{\mbox{\scriptsize ph}}_\mu$ as well as the background field ${\cal A}_\mu$ generated by the monopole-instantons.  As a check that our formalism gives the standard monopoles interaction,  we can turn off the photon field in (\ref{total lagrangian})  to find that $\frac{1}{4g_3^2}\int d^3x  F_{\mu\nu} F_{\mu\nu}=\frac{1}{4g_3^2}\int d^3x  {\cal F}_{\mu\nu} {\cal F}_{\mu\nu}=\frac{4\pi }{g_3^2 }\sum_{a >  b}\frac{q_aq_b}{| x_a- x_b|}$,  which is the monopole-monopole Coulomb interaction. In Section 4, it will be clear how to carry out the path integral rigorously by using an abelian duality transformation. 

The partition function (\ref{total partition function}) encodes all information about the system under study.
\footnote{However, this partition function does not contain information about the N-ality. For example, the N-ality zero sector should obey the perimeter rather than the area law. This is the main criticism of the abelian confinement mechanism, see \cite{Greensite:2011zz}.}
For example, as Polyakov did, one can completely ignore the W-bosons at zero temperature to find that there is a linear confining potential between two external charged probes.  At finite temperature $T$, we compactify the Euclidean time over a circle of radius $1/T$. At low temperatures compared to the W-boson mass $T<M_W$, one can integrate out the W-bosons. This results in an interacting gas  of W-bosons and magnetic monopoles. Since at low temperatures the W-bosons have non-relativistic speeds,  in the following we consider a non-relativistic version of the partition function (\ref{total partition function}). We will not try to directly start from (\ref{total partition function}) and take the non-relativistic limit. Instead, we will write down a partition function motivated by the physics of the problem. Since we have a system of  W-bosons and monopole-instantons, it is tempting to write down a many-body partition function of a non-relativitic gas of physical particles (W-bosons) in the background of  external field generated by an ensemble of monopole-instantons. Many-body treatment of a non-relativistic gas is a standard procedure in condensed matter that can be found in many books on the subject, see e.g. \cite{Wen:2004ym,mahan,Tsvelik}. The new thing here, which has not been considered before, is that we add instantons to the system as background field.

\subsection{The non-relativistic partition function}

Let us consider a two-dimensional gas (remember that we are working in $2+1$ dimensions) of interacting W-bosons of mass $M_W$ and charges $q_A=\pm 1$. These charged W-bosons experience logarithmic Coulomb interactions. In addition, let us consider this gas in the background of monopole-instantons which act as external time-dependent sources.  The classical Hamiltonian of the system reads
\begin{eqnarray}
\nonumber
H&=&\sum_{A}M_W+\sum_{A}\frac{(\vec p_A-eg_m q_A\vec {\cal A}(\vec x_A,x_0))^2}{2M_W}+i\sum_A  eg_mq_A{\cal A}_0(\vec x_A,x_0)\\
&&-\frac{e^2}{4\pi}\sum_{A \neq B} q_Aq_B\log|\vec x_A-\vec x_B|\,,
\label{non relativistic Hamiltonian}
\end{eqnarray} 
where $\vec x_A$ is the two-dimensional position of the W-boson, while $x_{0A}$ is its Euclidean time. The expressions for the monopole-instanton field, $\vec {\cal A}=({\cal A}_1,{\cal A}_2)$ and ${\cal A}_0$, are given by (\ref{monopole field}) and (\ref{total monopole field}). The first term in (\ref{non relativistic Hamiltonian}) is the rest mass of the W-boson gas. The second term is  the kinetic term of the non-relativistic  W-bosons written in terms of the kinetic momentum $\vec p_A-eg_m q_A\vec {\cal A}(\vec x_A,x_0)$. The third term describes the interaction between the W-bosons and the zeroth-component of the monopole-instantons field ${\cal A}_0$.  The factor $i$ is acquired because of working in the Euclidean space. In fact, this term is the Aharonov-Bohm coupling between the electrically charged W-bosons and the magnetically charged monopole-instantons. Finally, the last term is the mutual Coulomb interaction between two W-bosons. 

The second-quantized version of the Hamiltonian (\ref{non relativistic Hamiltonian}) can be obtained by replacing $\vec p \rightarrow -i\vec \nabla$, and introducing the $W^\pm$ density operators  $\hat\rho_{W_+}(\vec x,x_0)$ and $\hat\rho_{W_-}(\vec x,x_0)$ for the positively and negatively charged W-bosons, respectively:
\begin{eqnarray}
\nonumber
\rho_{W^+}(\vec x,x_0)&=&\sum_{A, q_A=+ 1}  \delta (\vec x-\vec x_A(x_0)) \rightarrow \hat\rho_{W^+}(\vec x,x_0)= \hat\Phi_+^\dagger(\vec x,x_0)\hat\Phi_+(\vec x,x_0)\,,\\
\rho_{W^-}(\vec x,x_0)&=&\sum_{A, q_A=- 1}  \delta (\vec x-\vec x_A(x_0)) \rightarrow \hat\rho_{W^-}(\vec x,x_0)= \hat\Phi_-^\dagger(\vec x,x_0)\hat\Phi_-(\vec x,x_0)\,.
\label{definition of rhoW}
\end{eqnarray}
The fields $\hat\Phi_\pm$ and $\hat\Phi_\pm^\dagger$ are the annihilation and creation operators for the gauge bosons $W^\pm$. They satisfy the equal time commutation relations
\begin{eqnarray}
\left[\hat\Phi_\pm(\vec x, x_0),\hat\Phi_\pm^\dagger(\vec y, x_0) \right]=\delta^{(2)}\left(\vec x-\vec y\right)\,.
\label{the definition or rhoM}
\end{eqnarray} 
The equal-time commutators of all other fields vanish. We also introduce the monopole density operator
\begin{eqnarray}
\rho_m(\vec x,x_0)=\sum_a q_a \delta^{(2)} \left(\vec x-\vec x_a\right)\delta(x_0-x_{0a})\,.
\label{monopole density operator}
\end{eqnarray}
Then, the field-theoretical version of the  Hamiltonian (\ref{non relativistic Hamiltonian}) reads
\begin{eqnarray}
\nonumber
\hat H&=&M_W\hat N_W-\frac{1}{2M_W}\int d^2x \hat\Phi_+^\dagger\left[\vec\nabla -ieg_m \int d^2x' dx_0'\vec A^m(\vec x-\vec x',x_0-x_0')\rho_m(\vec x',x_0') \right]^2\hat\Phi_+\\
\nonumber
&&-\frac{1}{2M_W}\int d^2x \hat\Phi_-^\dagger\left[\vec\nabla -ieg_m \int d^2x'dx_0'\vec A^m(\vec x-\vec x',x_0-x_0')\rho_m(\vec x',x_0') \right]^2\hat\Phi_-\\
\nonumber
&&-\frac{e^2}{4\pi}\int d^2 x\int d^2 x' dx_0' \left(\hat \rho_{W^+}(\vec x, x_0)-\hat \rho_{W^-}(\vec x, x_0)\right)\log |\vec x-\vec x'|\left(\hat \rho_{W^+}(\vec x', x_0)-\hat \rho_{W^-}(\vec x', x_0)\right)\\
&&+ieg_m \int d^2 x\int d^2 x'  dx_0' \left[\hat\rho_{W_+}(\vec x,x_0 )-\hat\rho_{W_-}(\vec x,x_0 )\right] A^m_0(\vec x-\vec x',x_0-x_0')\rho_m(\vec x',x_0')\,,
\label{second quantized Hamiltonian}
\end{eqnarray}
and $\hat N_W$ is the conserved W-boson number operator
\begin{eqnarray}
\hat N_W=\int d^2 x \left(\hat\Phi_+^\dagger\hat\Phi_++\hat\Phi_-^\dagger\hat\Phi_-\right)\,.
\end{eqnarray}
The term $m_W \hat N_W$ can be thought of as a chemical potential $-\mu \hat N_W$ added to the Hamiltonian, where $\mu=-M_W$ is the W-boson rest mass.  The second-quantized version of the Lagrangian can be obtained from the Hamiltonian (\ref{second quantized Hamiltonian}) by using the standard procedure and keeping in mind that we are working in the Euclidean space:
\begin{eqnarray}
\hat L=-\hat H- \int d^2 x\left( \hat\Phi_+^\dagger\partial_{x_0}\hat\Phi_++\hat\Phi_-^\dagger\partial_{x_0}\hat\Phi_- \right)\,.
\label{the non relativistic Lagrangian}
\end{eqnarray}

The finite temperature non-relativistic grand  partition function can be obtained in two steps. First, we regard the annihilation and creation operators, $\hat \Phi_\pm$ and $\hat \Phi_\pm^\dagger$, as classical complex fields, $\Phi_\pm$ and  $\Phi_\pm^*$, and perform the path integral over the various fields. Then, as we did in the case of relativistic partition function, we perform a sum over an arbitrary number of monopole-instantons located at positions $(\vec x_a,x_{0a})$, taking their Coulomb interaction into account. The finite temperature effects can be taken automatically into account by compactifying the Euclidean time  over a circle of circumference $\beta$, where $\beta$ is the inverse temperature, $\beta=1/T$. Then, we demand that the  fields satisfy periodic boundary conditions over the circle.
\footnote{This is the natural choice for bosonic fields. On the other hand, the natural choice for fermions is anti-periodic boundary conditions.}
Thus, the partition function reads
\begin{eqnarray}
\nonumber
Z_{\mbox{\scriptsize non-rel}}&=&\sum_{N_{m\pm}, q_a=\pm 1 }\frac{\xi_{m}^{N_{m+} + N_{m-}}}{N_{m+}! N_{m-}!}\left(\prod_a^{N_{m+} + N_{m-}} \int d^3x_a \right) \left[{\cal D}\Phi_+\right]_{\beta}  \left[{\cal D}\Phi_-\right]_\beta  \left[{\cal D}\Phi^*_+\right]_\beta  \left[{\cal D}\Phi^*_-\right]_\beta\\
&&\quad \quad \quad\quad\times \exp\left[\int_0^\beta dx_0 L_{\mbox{\scriptsize FNR}}\right] \,,
\label{non relativistic total partition function}
\end{eqnarray}
where $\xi_m$ is the monopole fugacity given by (\ref{the monopole fugacity}), and the subscript $\beta$, for example in $\left[{\cal D}\Phi_+\right]_{\beta} $, indicates that the fields must satisfy the periodic boundary condition $\Phi_+(\vec x, x_0)=\Phi_+(\vec x, x_0+\beta)$.  The full non-relativistic Lagrangian $L_{\mbox{\scriptsize TNR}}$ is the sum of the Lagrangian (\ref{the non relativistic Lagrangian}) and the monopole-monopole interaction term, taking into account the periodicity of the different quantities over the thermal circle:
\begin{eqnarray}
\nonumber
L_{\mbox{\scriptsize FNR}}&=&-2\pi g_m^2\int d^2x \int d^2 x' \int _0^\beta dx_0'\rho_m (x) \frac{1}{|x-x'|^{(p)}}\rho_m(x')\\
\nonumber
&-&\int d^2x \Phi_+^*\left\{M_W+\partial_{x_0}- \frac{\left[\vec\nabla -ieg_m \int d^2x' \int_0^\beta dx_0'\vec A^{m(p)}(\vec x-\vec x',x_0-x_0')\rho_m(\vec x',x_0') \right]^2}{2M_W}\right\}\Phi_+\\
\nonumber
&-&\int d^2x \Phi_-^*\left\{M_W+\partial_{x_0}-\frac{\left[\vec\nabla -ieg_m \int d^2x' \int_0^\beta dx_0'\vec A^{m(p)}(\vec x-\vec x',x_0-x_0')\rho_m(\vec x',x_0') \right]^2}{2M_W}\right\}\Phi_-\\
\nonumber
&+&\frac{e^2}{4\pi}\int d^2 x\int d^2 x' \int_0^\beta dx_0' \left( \rho_{W^+}(\vec x, x_0)-\rho_{W^-}(\vec x, x_0)\right)\log |\vec x-\vec x'|\left( \rho_{W^+}(\vec x', x_0)- \rho_{W^-}(\vec x', x_0)\right)\\
&-&ieg_m \int d^2 x\int d^2 x' \int_0^\beta dx_0' \left[\rho_{W_+}(\vec x,x_0 )-\rho_{W_-}(\vec x,x_0 )\right] A_{0}^{m(p)}(\vec x-\vec x',x_0-x_0')\rho_m(\vec x',x_0')\,.
\label{total non relativistic Lagrangian}
\end{eqnarray}
The propagator  $1/|x-x'|^{(p)}$ as well as the periodic monopole-instantons field $A^{m(p)}_{\mu}$ are obtained by summing an infinite number of image charges along the compact dimension
\begin{eqnarray}
\nonumber
\frac{1}{|x-x'|^{(p)}}&=&\sum_{n=-\infty}^{\infty}\frac{1}{\sqrt{\left(\vec x- \vec x'\right)^2+\left(x_0-x_0'+n\beta\right)^2}}\,,\\
A^{m(p)}_{\mu}&=&\sum_{n=-\infty}^\infty A^m_{\mu}\left(\vec x, x_0+n\beta \right)\,,
\end{eqnarray}
where the superscript $(p)$ indicates the periodicity of the quantity. Notice that we also have to take into account the spin degeneracy factor $S_W=2$ for the W-bosons. The partition function (\ref{non relativistic total partition function}) is one of the main results of the present work. Let us note that the steps of going from (\ref{non relativistic Hamiltonian}) to (\ref{non relativistic total partition function}) is a standard procedure in many-body physics. However, the inclusion of instantons as background fields is new, and to the best of our knowledge, has not been incorporated before in many-body treatments. At low temperatures $T<M_W$, and by neglecting any relativistic effects, the relativistic (\ref{total partition function}) and the non-relativistic (\ref{non relativistic total partition function}) partition functions contain the same information, and in principle one can use either of them to extract the physics.

At this point, one can split the Lagrangian (\ref{total non relativistic Lagrangian}) into two parts: a free Lagrangian $L_{\mbox{\scriptsize FNR 0}}$ and interacting part  $L_I$ such that $L_{\mbox{\scriptsize FNR }}=L_{\mbox{\scriptsize FNR 0}}+L_I$, where
\begin{eqnarray}
\nonumber
L_{\mbox{\scriptsize FNR 0}}=-\int d^2x\left\{ \Phi_+^*\left(M_W+\partial_{x_0}- \frac{\nabla ^2}{2M_W}\right)\Phi_+
+\Phi_-^*\left(M_W+\partial_{x_0}- \frac{\nabla ^2}{2M_W}\right)\Phi_-\right\}\,,\\
\end{eqnarray}
and the rest of $L_{\mbox{\scriptsize TNR }}$ is defined to be $L_I$. Then, performing perturbation analysis, one can expand the partition function (\ref{non relativistic total partition function}) as
\begin{eqnarray}
\nonumber
Z_{\mbox{\scriptsize non-rel}}&=&\sum_{N_{m\pm}, q_a=\pm 1 }\frac{\xi_{m}^{N_{m+} + N_{m-}}}{N_{m+}! N_{m-}!}\left(\prod_a^{N_{m+} + N_{m-}} \int d^3x_a\right)  \left[{\cal D}\Phi_+\right]_{\beta}  \left[{\cal D}\Phi_-\right]_\beta  \left[{\cal D}\Phi^*_+\right]_\beta  \left[{\cal D}\Phi^*_-\right]_\beta\\
&&\times \exp\left[\int_0^\beta dx_0 L_{\mbox{\scriptsize TNR 0}}\right]\sum_{n=0}^{\infty}\frac{1}{n!}\int_0^\beta dx_{01}...\int_0^\beta dx_{0n} L_{I}(x_{01})...L_{I}(x_{0n})\,.
\label{expansion in non relativistic partition function}
\end{eqnarray}
 The expansion (\ref{expansion in non relativistic partition function}) makes sense only if there is a small expansion parameter. In case of W-W interaction, the small parameter is taken to be the charge $e$. However, for the Aharonov-Bohm term we have $eg_m=1$. In this case, the true expansion parameter is the small monopole density $\rho_m$ which is a prerequisite for the validity of the monopole-instanton dilute gas approximation. 

In the next section, we use the partition function (\ref{expansion in non relativistic partition function}) to calculate the string tension between two external electric probes. In the absence of monopole-instantons the potential between the two probes is logarithmic. However, as Polyakov showed long time ago \cite{Polyakov:1976fu}, including the instantons in the background creates a mass gap ${\cal M}$ in the system, which in turn changes the logarithmic behavior into a linear confining potential between the probs. One expects the change from a logarithmic to linear behavior to happen at distances~${\cal M}^{-1}$. However, it was never shown explicitly how this happens. Using the above formalism, we show this smooth transition takes place, as expected, at distance of order of the inverse mass gap.

\section{The  potential between two external electric probes at $T=0$}

\subsection{An effective partition function and the Polyakov loop correlator }

In this section, we use the perturbative expansion of the partition function (\ref{expansion in non relativistic partition function}) to calculate the potential between two external electric charges located  in the background of the monopole-instanton gas, which is the same as calculating the Polyakov loop (electric) correlator. We will perform our analysis at zero temperature, or in other words, at infinite compactification radius $\beta \rightarrow \infty$. However, we retain all expressions as a function of $\beta$ such that the limit $\beta \rightarrow \infty$ should be understood implicitly.  At temperatures lower than the W-boson mass,  the contribution coming from the W-bosons is accompanied by a Boltzamann suppression factor $e^{-M_W/T}$. Thus, the dynamics of the W-bosons is completely suppressed at $T=0$, and one can neglect the second, third, fourth, and last terms in the Lagrangian (\ref{total non relativistic Lagrangian}). Then, we are left only with the first term, the monopole-monopole interaction. Now, let us introduce two external probes with electric charges $Ze$ and $-Ze$ located at $\vec R_1$ and $\vec R_2$, respectively. We take these charges to be infinitely massive. Hence, the free Hamiltonian takes the form
\begin{eqnarray}
H_0=\frac{Z^2 e^2 }{2\pi} \log|\vec R_1-\vec R_2|\,.
\end{eqnarray}
 These prob charges also see a distribution of monopole-instantons in the background, and couple to them via the Aharonov-Bohm  interaction term similar to the last term in (\ref{total non relativistic Lagrangian}). Therefore, the interaction Hamiltonian reads
\begin{eqnarray}
\nonumber
&&2\pi g_m^2\int d^2x \int d^2 x' \int _0^\beta dx_0'\rho_m (x) \frac{1}{|x-x'|^{(p)}}\rho_m(x')\\
\nonumber
&&+ieg_m\int d^2 x \int d^2 x' \int_0^\beta dx_0' \left[Z \delta^{(2)}(\vec x-\vec R_1)-Z\delta^{(2)}(\vec x-\vec R_2) \right] A^{m(p)}_{0}(\vec x-\vec x',x_0-x_0')\rho_m(\vec x', x_0')\,,\\
\label{interaction Hamiltonian for electric PF}
\end{eqnarray}
where the Dirac-delta functions, $\delta^{(2)}(\vec x-\vec R_1)$ and $\delta^{(2)}(\vec x-\vec R_2)$, amount to placing the infinitely heavy  charges at $\vec R_1$ and $\vec R_2$.

Before proceeding, let us elucidate the physics of the second line in (\ref{interaction Hamiltonian for electric PF}). The corresponding action can be written as
\begin{eqnarray}
\nonumber
S_{\mbox{\scriptsize 2nd line}}&=&-iZeg_m\int d^2 x' \int_0^\beta dx_0'\rho_m(\vec x', x_0')\int_0^\beta dx_0 \left[ A^{m(p)}_{0}(\vec R_1-\vec x',x_0-x_0')\right.\\
&&\left.\quad\quad\quad\quad - A^{m(p)}_{0}(\vec R_2-\vec x',x_0-x_0') \right]\,.
\label{the physics of the second line}
\end{eqnarray}
The integral $\int_0^\beta dx_0 A^{m(p)}_{0}(\vec x,x_0)$ can be done exactly
\begin{eqnarray}
\int_0^\beta dx_0 A^{m(p)}_{0}(\vec x,x_0)&=&\int_0^\beta dx_0 \sum_{n=-\infty}^{\infty} A_0^m (\vec x,x_0+n\beta)=\int_{-\infty}^\infty dx_0 A_0^m(\vec x, x_0)=2\Theta(\vec x)\,,
\label{manipulating the monopole field}
\end{eqnarray}
where 
\begin{eqnarray}
\Theta(\vec x)=-\mbox{sign}(x_1)\frac{\pi}{2}+\mbox{Arctan}\frac{x_2}{x_1}\,.
\label{the angle theta}
\end{eqnarray} 
The angle $\Theta(\vec x)$ is a purely two-dimensional quantity that has a discontinuity on the negative $x_2$-axis. This should be expected since in the gauge (\ref{monopole field}) the Dirac string stems from the monopole-instanton and runs along the negative $x_2$-axis. Thus, one can rewrite (\ref{the physics of the second line}) in the form
\begin{eqnarray}
S_{\mbox{\scriptsize 2nd line}}&=&-iZeg_m\int d^2 x' \int_0^\beta dx_0'\rho_m(\vec x', x_0')\left\{2\Theta\left(\vec R_1-\vec x'\right)-2\Theta\left(\vec R_2-\vec x'\right) \right\}\,.
\label{the physics of the second line in short form}
\end{eqnarray} 
To understand the physics of (\ref{the physics of the second line in short form}), we compare the expression between braces to the expectation value of the Polyakov loop correlator $\langle e^{i\oint_{C=S^1\times S^1} dx_\mu A_\mu^m}\rangle$, where the two circles wind around the thermal direction. This correlator gives us information about the potential between two infinitely heavy electric probes.  To understand the geometry of the loop $C$, we start by considering a rectangular loop that lies on the $x_0-x_2$ plane at $x_1=0$. The Polyakov loop correlator is a gauge invariant quantity and in principle one can choose the loop to lie on any plane. However, our many-body Hamiltonian (\ref{interaction Hamiltonian for electric PF}) is written in terms of the potential $A_0^{m(p)}$ which is a gauge-dependent quantity. Therefore, given the gauge (\ref{monopole field}), one is forced to place the loop in a specific plane (the $x_0-x_2$ plane in our case) in order to hide the Dirac string singularities, as we will demonstrate shortly. The loop $C$ extends from $x_2=R_1$ (the position of the first probe) to $x_2=R_2$ (the position of the second probe), and from $x_0=-\beta/2$ to $x_0=\beta/2$. Because the $x_0$ coordinate has the geometry of a circle, we identify the edges $-\beta/2$ and $\beta/2$ keeping in mind that we are working in the limit $\beta \rightarrow \infty$. This results in the loop $C=S_{1\,\, \mbox{\scriptsize at}\,\, x_2=R_1}\times S_{1\,\, \mbox{\scriptsize at}\,\, x_2=R_2}$ or $C=S_1\times S_1$ for short. Then, using Gauss's theorem, we obtain  $ \langle e^{i\int_{S^1\times S^1}dx_\mu A_\mu^m}\rangle= \langle e^{i\int_{C_2}dS_\mu B_\mu^m}\rangle$, where $\int_{C_2}dS_\mu B_\mu^m$ is the magnetic flux through the cylinder $C_2$ wrapping the thermal direction. This geometry is illustrated in Figure (\ref{cylinder}).  The magnetic field due to magnetic charge density $\rho_m$ is $B^m_\mu(x)=\int d^3 x' \rho_m(\vec x', x_0')\left(\frac{ (x-x')_\mu}{|x-x'|^3}\right)^{(p)}$. Hence, the quantity $\int_{C_2} dS_\mu B_\mu^m$ can be written as $\int d^3 x' \eta(\vec x',x_0')\rho_m(\vec x',x_0')$, where $\eta(\vec x',x_0')=\int_{C_2} dS_\mu\left(\frac{(x'-x)_\mu}{|x'-x|^3}\right)^{(p)}$ is the magnetic flux through the surface $C_2$ due to a unit magnetic charge located at $(\vec x',x_0')$. The computation of $\eta(\vec x',x_0')$ is carried out in Appendix A to find $\eta(\vec x',x_0')=2\Theta\left(\vec R_1-\vec x'\right)-2\Theta\left(\vec R_2-\vec x'\right)$, which is the quantity that appears in (\ref{the physics of the second line in short form}).  Further, by studying the properties of $\eta$, we find that the appropriate jump in the flux happens when the two electric probes are positioned along the $x_2$-axis. This is illustrated in Figure (\ref{the string position}) which explains why we placed the loop in the $x_0-x_2$ plane. The only place where we can allow for a  discontinuity in the flux is across the line connecting the two charges. Had we placed the charges along any other axis, the flux would be discontinuous across arbitrary places.
 \begin{FIGURE}[ht]
    {
    \parbox[c]{\textwidth}
        {
        \begin{center}
        \includegraphics[angle=0, scale=.8]{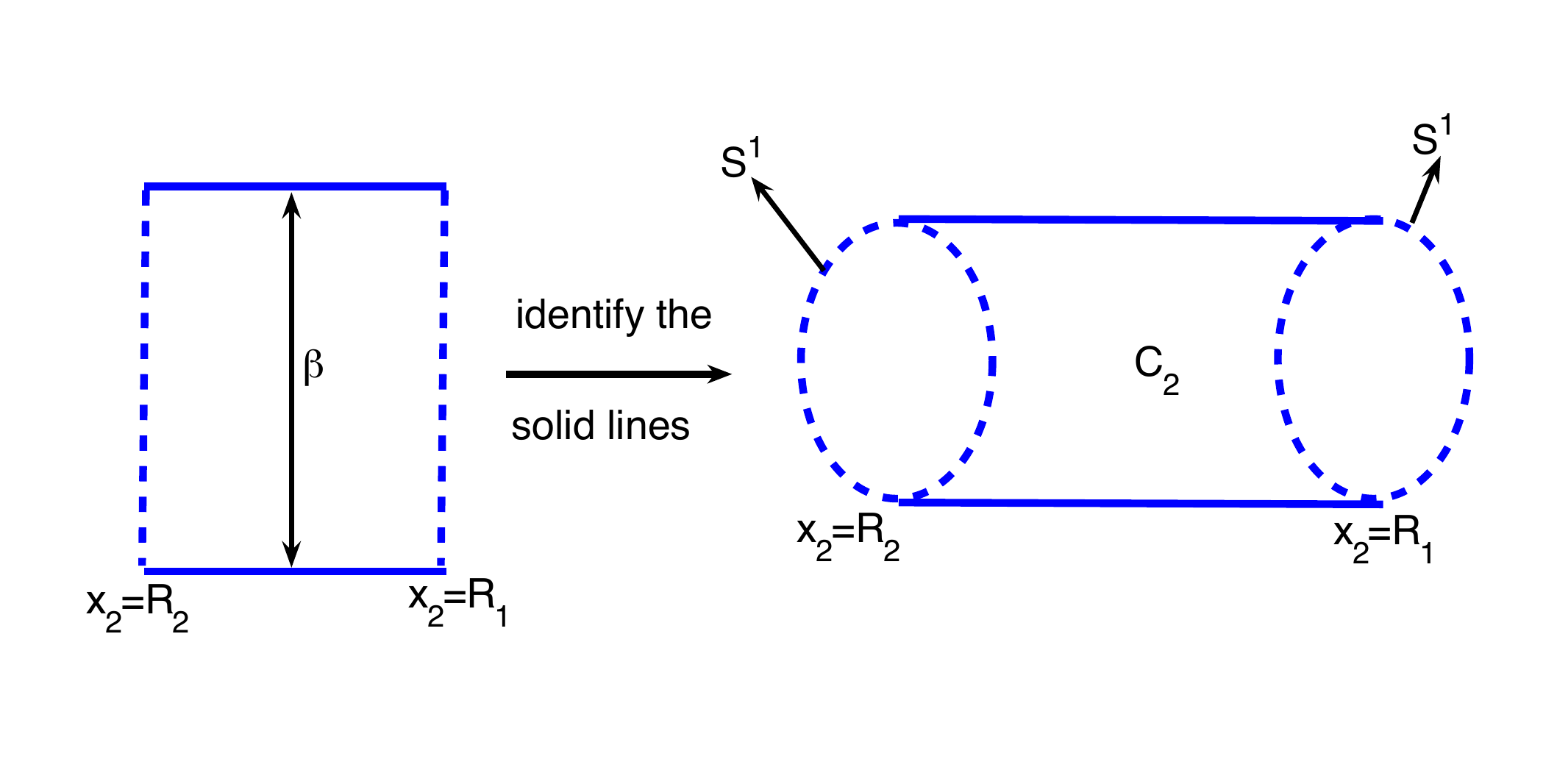}
	\hfil
        \caption 
      {The geometry of the Wilson Loop. 
			}
      \label{cylinder}
       \end{center}
        }
    }
\end{FIGURE}
 \begin{FIGURE}[ht]
    {
    \parbox[c]{\textwidth}
        {
        \begin{center}
        \includegraphics[angle=0, scale=.7]{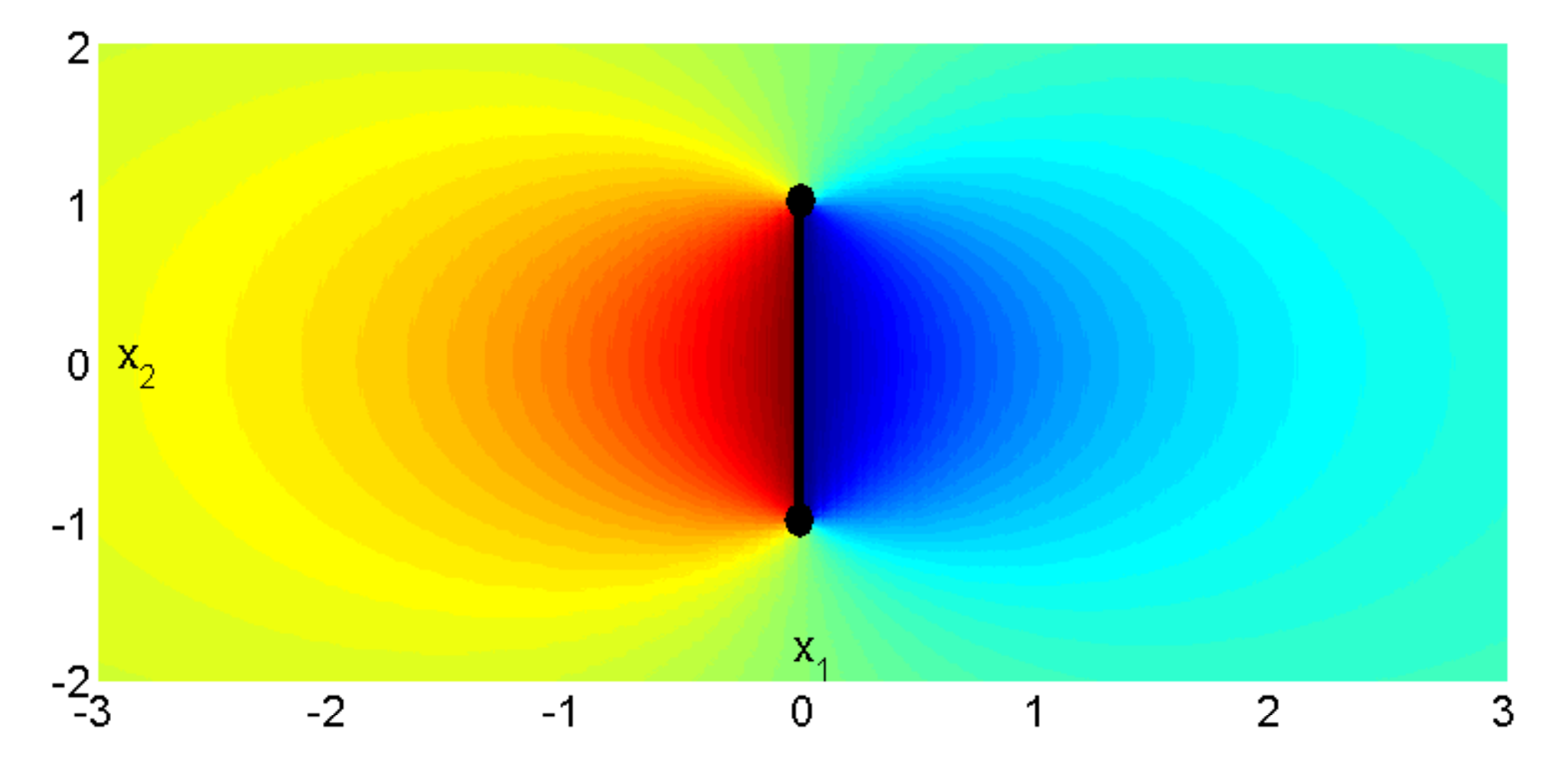}
	\hfil
        \caption 
      {A topographical plot of the flux $2\Theta\left(\vec R_1-\vec x'\right)-2\Theta\left(\vec R_2-\vec x'\right)$. The electric probes are located at $(0,1)$ and $(0,-1)$. We see that there is a discontinuity in the flux  along the $x_2$-axis which is bounded by the location of the two probes.
			}
      \label{the string position}
       \end{center}
        }
    }
\end{FIGURE}

 In order to further simplify our analysis,  we use a mean-field approach to replace the  monopole-monopole Coulomb interaction (the first line in (\ref{interaction Hamiltonian for electric PF})) with an effective description. In this approach, one does not have to account for the individual behavior of each monopole. Rather, we use the fact that the monopole-instantons form a plasma of positive and negative charges  to deduce the form of the collective behavior of the monopole density correlation function $\left\langle \rho_m(\vec x, x_0) \rho_m(0) \right\rangle$, where the brackets $\left\langle\quad\right\rangle$ denote a statistical average. We  explicitly compute this quantity in the next section assuming the fluctuations of the density to follow a Gaussian distribution, i.e. $\left\langle \rho_m(\vec x, x_0)\right\rangle=0$. Therefore, one can replace the full partition function (\ref{expansion in non relativistic partition function}), which involves a sum over an arbitrary number of monopole-instantons, with the "effective" partition function
\begin{eqnarray}
{\cal Z}_e=e^{-\beta\frac{Z^2 e^2 }{2\pi} \log|\vec R_1-\vec R_2|}\sum_{n=0}^\infty\frac{1}{n!}\int_0^\beta dx_{01}...\int_0^\beta dx_{0n} \left\langle L_I(x_{01})L_I(x_{02})...L_I(x_{0n})  \right\rangle\,,
\label{effective partition function}
\end{eqnarray}
where
\begin{eqnarray}
\nonumber
L_I(x_0)=-ieg_m\int d^2 x \int d^2 x' \int_0^\beta dx_0' \left[Z \delta(\vec x-\vec R_1)-Z\delta(\vec x-\vec R_2) \right] A_{0}^{m(p)}(\vec x-\vec x',x_0-x_0')\rho_m(\vec x', x_0')\,.\\
\label{interaction potential in the effective description }
\end{eqnarray}
The subscript in ${\cal Z}_e$ denotes the effective partition function of the "electric" probes. This is to distinguish it from the effective partition function of the "magnetic" probes that we consider in the next section.  

Since the statistical average in (\ref{interaction potential in the effective description }) is performed over Gaussian fluctuations, we have for any odd correlator $\left\langle L_I( x_{01})...L_I(x_{0 2n+1})\right\rangle=0$. On the other hand, one can use the Wick's theorem to write the even correlators in terms of the lowest moment  $\left\langle L_I(x_{01}) L_I(x_{02}) \right\rangle$:
\begin{eqnarray}
\nonumber
&&\int_0^\beta dx_{01}...\int_0^\beta dx_{02n}\left\langle L_I(x_{01})L_I(x_{02})...L_I(x_{02n})  \right\rangle=\\
&&\quad\quad\quad\quad\frac{(2n)!}{2^{n}n!}\left[\int_0^\beta dx_{01}\int_0^\beta dx_{02}\left\langle L_I(x_{01})L_I(x_{02})\right\rangle\right]^n\,.
\label{even correlator}
\end{eqnarray}
Substituting (\ref{even correlator}) into (\ref{effective partition function}), we finally obtain
\begin{eqnarray}
{\cal Z}_e=\exp\left[-\beta\frac{Z^2 e^2 }{2\pi} \log|\vec R_1-\vec R_2|+\frac{1}{2}\int_0^\beta dx_{01}\int_0^\beta dx_{02}\left\langle L_I(x_{01})L_I(x_{02})\right\rangle \right]\,.
\end{eqnarray}
To calculate the correlator $\left\langle L_I(x_{01})L_I(x_{02})\right\rangle$, it is more appropriate to go to the Fourier basis. The Fourier decomposition of $A_0^{m(p)}$ and $\left\langle \rho_m(\vec x_1, x_{01})\rho_m(\vec x_2, x_{02})\right\rangle$ is given by
\begin{eqnarray}
\nonumber
A_0^{m(p)}(\vec x,x_0)&=&\frac{1}{\beta}\int \frac{d^2p}{\left(2\pi\right)^2}\sum_{m=-\infty}^{\infty} e^{-ip_mx_0}e^{-i\vec p\cdot \vec x}\tilde A_0^{m(p)}(\vec p, p_m)\,,\\
\langle \rho_m(\vec x_1,x_{01})\rho_m(\vec x_2,x_{02}) \rangle&=&\frac{1}{\beta}\int \frac{d^2 p}{\left(2\pi\right)^2}\sum_{m=-\infty}^{\infty}e^{-ip_m(x_{01}-x_{02})}e^{-i\vec p \cdot(\vec x_1-\vec x_2)}\Pi(\vec p,p_m)\,,
\label{Fourier decomposition}
\end{eqnarray}
where $p_m=2\pi m/\beta$ are the Matsubara frequencies.
\footnote{In general, given the periodic function $F^{(p)}(\vec x, x_0)$, the Fourier transform reads
\begin{eqnarray}
F^{(p)}(\vec x, x_0)=\frac{1}{\beta}\int \frac{d^2p}{\left(2\pi\right)^2}\sum_{m=-\infty}^{\infty} e^{-ip_mx_0}e^{-i\vec p\cdot \vec x}\tilde F^{(p)}(\vec p, p_m)\,,
\end{eqnarray}
while the inverse Fourier transform is given by
\begin{eqnarray}
\tilde F^{(p)}(\vec p, p_m)=\int d^2 x \int _{0}^\beta dx_0 e^{ip_mx_0}e^{i\vec p\cdot \vec x}F^{(p)}(\vec x, x_0)\,.
\label{the Fourier transformed F}
\end{eqnarray}
}
The form of the Fourier decomposition of the density-density correlation function, such that $\Pi$ depends only on a single momentum $(\vec p, p_m)$, is a consequence of the fact that the monopole-instanton plasma is invariant under spacetime translation. Substituting (\ref{Fourier decomposition}) into $\left\langle L_I(x_{01})L_I(x_{02})\right\rangle$ we find
\begin{eqnarray}
\nonumber
{\cal Z}_e&=&\exp\left[-\beta\frac{Z^2 e^2 }{2\pi} \log|\vec R_1-\vec R_2|\right.\\
\nonumber
&&\left.+\beta Z^2e^2g_m^2\int \frac{d^2 p}{\left(2\pi\right)^2}\left[ e^{-i\vec p\cdot (\vec R_1-\vec R_2)}-1\right]\tilde A_0^{m(p)}(\vec p,0)\tilde A_0^{m(p)}(-\vec p,0)\Pi(\vec p,0)\right]\,.\\
\label{final effective partition function for the Polyakov loop}
\end{eqnarray}
Since the external probes are not dynamical, we notice that the expression above has a contribution coming only from the zero Matsubara frequency. The second term in (\ref{final effective partition function for the Polyakov loop}) can be represented diagrammatically as in Figure (\ref{The Feynman diagram}). This expression can also be obtained by performing a perturbative expansion to the expectation value of the Wilson loop, which is presented in Appendix A.   Before proceeding to calculate the potential between the external probes in the background of the monopole-instanton plasma, we pause to calculate the correlation function $\Pi(\vec p,0)$ using a mean-field approach.  
 \begin{FIGURE}[ht]
    {
    \parbox[c]{\textwidth}
        {
        \begin{center}
        \includegraphics[angle=0, scale=.8]{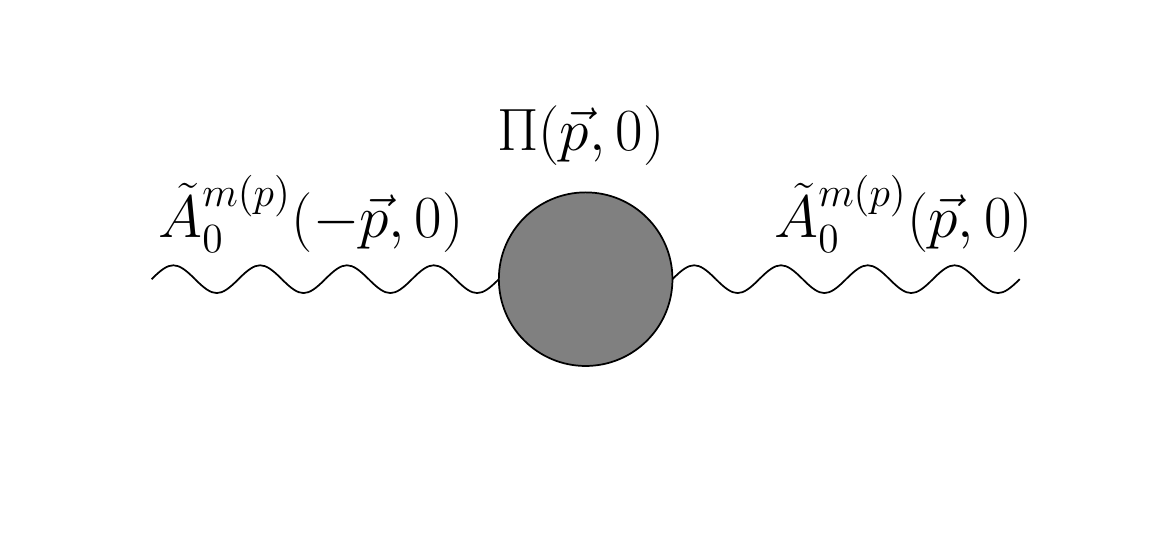}
	\hfil
        \caption 
      {A Feynman graph representation of the second line in (\ref{final effective partition function for the Polyakov loop}). 
			}
      \label{The Feynman diagram}
       \end{center}
        }
    }
\end{FIGURE}
%

\subsection{The monopole-instanton density-density correlation function: the 't Hooft loop correlator}

The density-density correlation function of the monopole-instanton plasma can be obtained in a similar fashion to what we did in the previous section. In this regard, we introduce two infinitely heavy external magnetic probes with magnetic charges $Zg_m$ and $-Zg_m$ located respectively at $R_1$ and $R_2$, where $R_1$ and $R_2$ are the three-dimensional Euclidean position vectors. Calculating the effective potential between these two external probes amounts to the computation of the 't Hooft magnetic correlator.  Since these calculations are performed at $T=0$, we drop the contribution from the W-bosons. The free action due to the interaction between the two probes reads
\begin{eqnarray}
S_0= -4 \pi Z^2 g_m^2\frac{1}{|x-x'|^{(p)}}\,.
\end{eqnarray}
The interaction Hamiltonian of the monopole gas in the presence of the external monopole-instanton impurities is 
\begin{eqnarray}
\nonumber
&&2\pi g_m^2\int d^2x \int d^2 x' \int _0^\beta dx_0'\rho_m (x) \frac{1}{|x-x'|^{(p)}}\rho_m(x')\\
&&\quad\quad\quad+4\pi g_m^2\int d^2x d^2 x' \int _0^\beta dx_0'\left[Z\delta(\vec x-R_1)-Z\delta(x-R_2) \right]\frac{1}{|x-x'|^{(p)}}\rho_m(x')\,.
\label{interaction Hamiltonian for the instantons gas}
\end{eqnarray}
As we did before, we omit the first term in (\ref{interaction Hamiltonian for the instantons gas}) and replace it  by an effective description. In this description we encode the effect of the monopole-instanton plasma in the density-density correlation function. Again, we assume that the fluctuations in the density function follow a Gaussian distribution, i.e. $\left\langle \rho_m(\vec x, x_0)\right\rangle=0$.  By repeating the same steps that led us from (\ref{interaction Hamiltonian for electric PF}) to (\ref{final effective partition function for the Polyakov loop}) we obtain the partition function that describes the "effective" monopole-instanton plasma in the presence of the external magnetic probes: 
\begin{eqnarray}
\nonumber
{\cal Z}_m&=&\exp\left[\frac{-Z^2g_m^2}{\beta}\int \frac{d^2 p}{\left(2\pi\right)^2}\sum_{m=-\infty}^\infty e^{-ip_m(R_{01}-R_{02})}e^{-i\vec p\cdot (\vec R_1-\vec R_2)}\tilde V^{(p)}(\vec p,ip_m)\right.\\
&&\left.\quad\quad\quad\quad\quad\quad\quad\times\left[1-g_m^2\tilde V^{(p)}(-\vec p,-ip_m)\Pi(\vec p, ip_m)\right]\right]\,,
\label{magnetic partition function}
\end{eqnarray}
where the subscript in ${\cal Z}_m$ denotes the effective "magnetic" partition function. According to (\ref{the Fourier transformed F}),  the potential $\tilde V^{(p)}(\vec p,ip_m)$ is given by
\begin{eqnarray}
\nonumber
\tilde V^{(p)}(\vec p, ip_m)&=&\int d^2x \int_0^\beta dx_0 e^{ip_mx_0} e^{i\vec p \cdot \vec x}\frac{4\pi}{|x|^{(p)}}=\int d^2x \int_0^\beta dx_0 e^{ip_mx_0} e^{i\vec p \cdot \vec x}\sum_{n=-\infty}^{\infty}\frac{4\pi}{\sqrt{\vec x^2+(x_0+n\beta)^2}}\\
&=&\int d^2 x\int_{-\infty}^{\infty} dx_0 e^{ip_mx_0} e^{i\vec p \cdot \vec x}\frac{4\pi}{\sqrt{\vec x^2+x_0^2}}=\frac{16\pi^2}{ p^2+p_m^2}\,.
\end{eqnarray}

The quantity between brackets in (\ref{magnetic partition function}) can be recognized as an effective dielectric constant for the monopole-instanton plasma:
\begin{eqnarray}\label{ dielectric constant}
\frac{1}{\epsilon(\vec p,ip_m)}=1-g_m^2\tilde V^{(p)}(-\vec p,-ip_m)\Pi(\vec p, ip_m)\,.
\end{eqnarray}
This dielectric constant can be modeled using a mean-field approach. The physics of the three-dimensional monopole-instanton plasma is that the Coulomb interactions are screened. Thus, the potential between two external probes is given by
\begin{eqnarray}
\tilde V^{(p)}_s(\vec p, ip_m)=\frac{16\pi^2}{p^2+p_m^2+{\cal M}^2}\,,
\end{eqnarray}
where ${\cal M}$ is the mass gap or equivalently the inverse Debye screening length of the monopole plasma. 
\footnote{ Contrasting the form of the density-density correlation $\Pi(\vec p, p_m)$ obtained from the mean-field approach with the same quantity obtained using the Polyakov calculations (reviewed in Appendix A), one can read the relation between the mass gap and the monopole fugacity 
\begin{eqnarray}
{\cal M}^2=\mbox{constant} \times M_W^5 g_3^{-6}\exp\left[-4\pi \frac{M_W}{g_3^2}\epsilon\left(\frac{M_W}{m_H}\right)\right]\,.
\end{eqnarray}
}
 This potential can be written as $\tilde V^{(p)}_s(\vec p, ip_m)=16\pi^2/[(p^2+p_m^2)\epsilon(p,ip_m)]$, where $\epsilon(p,ip_m)=1+{\cal M}^2/(p^2+p_m^2)$. Using the definition (\ref{ dielectric constant}) we obtain 
\begin{eqnarray}
\label{effective density}
\Pi(\vec p, p_m)=\frac{1}{16\pi^2 g_m^2}\frac{\left(p^2+p_m^2\right) {\cal M}^2}{p^2+p_m^2+{\cal M}^2}\,.
\label{the definition of PI}
\end{eqnarray}
%

\subsection{Calculations of the confinement potential}

Having calculated the density-density correlator, the final step before applying the master formula (\ref{final effective partition function for the Polyakov loop}) is to calculate the Fourier transform of $A^{m(p)}(\vec x, x_0)$:
\begin{eqnarray}
\nonumber
\tilde A^{m(p)}_0(\vec p, 0)&=&\int d^2 x \int_0^\beta dx_0 e^{i\vec p \cdot \vec x}\sum_{n=-\infty}^{\infty}A_0^m(\vec x, x_0+n\beta)=\int d^2 x \int_{-\infty}^{\infty} dx_0 e^{i\vec p \cdot \vec x}A_0^m(\vec x, x_0)\\
&=&2\int d^2 x  e^{i\vec p \cdot \vec x}\Theta(\vec x)\,,
\label{Fourier transform of A0}
\end{eqnarray}
where $\Theta (\vec x)$ is given by (\ref{the angle theta}). The remaining two-dimensional integral can be done by using methods of generalized Fourier transform. Alternatively, we proceed by recalling the discussion after (\ref{the angle theta}). There, we argued that there is a discontinuity in the magnetic flux as we cross the $x_1$-axis, while the flux is continuous along the $x_2$-axis. Then, we use the two-dimensional duality between the angle and logarithm $\partial_i\theta (\vec x)=\epsilon_{ij}\partial_j\log|\vec x|$, where $\epsilon_{12}=-1$. Since $\theta$ is continuous along the $x_2$-axis, we demand that $\partial_2\theta(\vec x)=\partial_1 \log |\vec x|$. The Fourier transform of the previous relation is $p_2\theta(\vec p)=p_1\int d^2 x e^{i\vec p\cdot x}\log |\vec x|=-2\pi \frac{p_1}{p^2}$. Then, we obtain 
\begin{eqnarray}
\tilde A^{m(p)}_0(\vec p, 0)=-\frac{4\pi p_1}{p_2(p_1^2+p_2^2)}\,.
\end{eqnarray}

Finally, we are in a position to calculate the potential between two external probes. Putting everything together and remembering that the electric probes are separated a distance $R$ along the $x_2$-axis, we find
\begin{eqnarray}
\nonumber
-T\log{\cal Z}_e&=& \frac{Z^2e^2}{2\pi}\log R+Z^2e^2 {\cal M}^2\int \frac{d^2p}{\left(2\pi\right)^2} \frac{\left(1- e^{-i p_2R}\right)p_2^2}{p_1^2\left(p_1^2+p_2^2\right)\left(p_1^2+p_2^2+{\cal M}^2\right)}\\
&=&\frac{Z^2e^2}{2\pi}\left[\log R +\int_0^\infty dx \frac{1-\cos\left(x{\cal M}R\right)}{x^2\left(x+\sqrt{x^2+1}\right)} \right]\,.
\end{eqnarray} 
The integral can be obtained in a closed form in terms of the Meijer-$G$ function \cite{Bateman}, and the total potential between the electric probes read
\begin{eqnarray}
V(R)=\frac{Z^2e^2}{2\pi}\left(\log R+\frac{1}{4}G^{2,3}_{3,5}\left[\left.\begin{array}{ccccc} 1,&1,&\frac{3}{2}& & \\1, & 1,&0,&0,&\frac{1}{2} \end{array}\right|\frac{{\cal M}^2R^2}{4} \right] \right)\,.
\label{The potential in terms of Meijer G function}
\end{eqnarray}
This is one of the main results in the present work. For $R{\cal M}>>1$, we use the asymptotic expansion
\begin{eqnarray}
\frac{1}{4}G^{2,3}_{3,5}\left[\left.\begin{array}{ccccc} 1,&1,&\frac{3}{2}& & \\1, & 1,&0,&0,&\frac{1}{2} \end{array}\right|\frac{{\cal M}^2R^2}{4} \right]\stackrel{R{\cal M}>>1}{\longrightarrow}\frac{\pi {\cal M}R}{2}-\log{\cal M}R-\sqrt{\frac{\pi}{2}}\frac{e^{-{\cal M}R}}{\left(R {\cal M}\right)^{3/2}}
\end{eqnarray}
to find that the logarithms cancel at large distances
\begin{eqnarray}
V(R)|_{{\cal M}R>>1}=\frac{Z^2e^2}{2\pi}\left(\frac{\pi {\cal M}R}{2}-\sqrt{\frac{\pi}{2}}\frac{e^{-{\cal M}R}}{\left(R {\cal M}\right)^{3/2}}\right)\rightarrow Z^2 \sigma_{\mbox{\scriptsize mean-field}}R\,,
\end{eqnarray}
where
\begin{eqnarray}
\sigma_{\mbox{\scriptsize mean-field}}=\frac{e^2{\cal M}}{4}
\label{the string tension using the mean field approximation}
\end{eqnarray}
is the string tension calculated using the mean-field approach. The behavior of $V(R)$ and its derivative (the electric field) is depicted in Figure (\ref{The potential and electric field profile between two external probes}). Interestingly enough, the smooth transition from logarithmic to linear potential is an indication of the conservation of the electric flux. Comparing $\sigma_{\mbox{\scriptsize mean-field}}$ to the string tension calculated originally in \cite{Polyakov:1976fu}, $\sigma_{\mbox{\scriptsize string-exact-Polyakov}}=g_3^2{\cal M}/(2\pi^2)$, we find that the mean-field value is off by a factor of $2/\pi^2$. Such discrepancy is attributed to strong coupling physics that is not taken care of in the mean-field approach. Another method to calculate $\sigma_{\mbox{\scriptsize mean-field}}$ is presented in Appendix A.

\section{The finite temperature effects}

In the previous section, we performed our calculations strictly at zero temperature. In this section, we consider the finite temperature effects. Unlike the zero temperature case, the W-bosons are now excited and will modify the confinement picture. Their effect is well pronounced near the confinement-deconfinement transition region where they play a prominent role. In order to take these effects into consideration, one needs a systematic approach to start from the relativistic partition function (\ref{total partition function}) and integrate out the W-bosons. This is achieved using a heat kernel expansion technique that takes into account the presence of the thermal holonomy. One can also obtain the same results starting from the non-relativistic partition function (\ref{non relativistic total partition function}). Such calculations work as a non-trivial check on the many-body approach described in the previous section.

\subsection{Integrating out the heavy fields: the effective action}

Our starting point is the relativistic partition function (\ref{total partition function}). This partition function encodes all information about the system at all temperatures. Basically, it includes a path sum over the monopole-instanton gas as well as the fluctuations of the electromagnetic $F^{\mu\nu}$, W-boson $W_{\mu}^{\pm}$, Higgs  $\phi$, Goldston boson $\phi^\pm$, and the ghost $c^\pm$ fields. It is important to emphasize again, as mentioned in section (2), that the electromagnetic field $F^{\mu\nu}$ includes both the photon $F_{\mu\nu}^{\mbox{\scriptsize ph}}$ and monopole-instanton ${\cal F}_{\mu\nu}$ contribution ; thus $F_{\mu\nu}=F_{\mu\nu}^{\mbox{\scriptsize ph}}+{\cal F}_{\mu\nu}$.  The fields $W_\mu^\pm$, $\phi^\pm$, and $c^\pm$  are massive with mass $M_W$, while the mass of the Higgs field depends on the quartic coupling constant $\lambda$. In the following, we are interested in the system behavior at temperatures lower than the W-boson mass, $T<M_W$. Hence, one can integrate out the fields $W_\mu^\pm$, $\phi^\pm$, and $c^\pm$. Since the Higgs field is massive, it is short ranged and does not participate in the dynamics. Therefore, we can equally well integrate it out or just leave it aside.

Integrating out the heavy fields $W_\mu^\pm$, $\phi^\pm$, $c^\pm$, and $\phi$, and ignoring the non-quadratic terms in (\ref{total lagrangian}), which contribute only to higher order effects, we obtain
\begin{eqnarray}
\nonumber
{\cal Z}_{\mbox{\scriptsize grand}}=&&\sum_{N_{m\pm}, q_a=\pm 1 }\frac{\xi_m^{N_{m+} + N{_m-}}}{N_{m+}! N_{m-}!}\left(\prod_a^{N_{m+} + N_{m-}} \int d^{d+1}x_a\right)  \\
\nonumber
&&\times\int [{\cal D} A^{\mbox{\scriptsize ph}}_\mu]\exp\left[-\int d^{d+1}x\frac{1}{4g_3^2} F_{\mu\nu} F_{\mu\nu}\right]\mbox{Det}\left[- D^2\delta_{\mu\nu}+{\cal M}_{\mu\nu}(x)  \right]^{-1}_{W_{\mu}^\pm}\\
&&\times\mbox{Det}\left[- D^2+M_W^2\right]^{-1}_{\phi^\pm} \mbox{Det}\left[- D^2+M_W^2\right]^2_{c^\pm}\mbox{Det}\left[-\partial^2+M_H^2\right]^{-1/2}_{\phi}\,,
\end{eqnarray}
where ${\cal M}_{\mu\nu}=2F_{\mu\nu}+\delta_{\mu\nu}M_W^2$. We leave the dimension of the spacetime unspecified and equal to $d+1$ setting $d=2$ at the end of calculations. Now, we use the identity $\mbox{Det} A=\exp[\mbox{Tr}\log A]$ to write the partition function as
\begin{eqnarray}
{\cal Z}_{\mbox{\scriptsize grand}}&=&\sum_{N_{m\pm}, q_a=\pm 1 }\frac{\xi_m^{N_{m+} + N_{m-}}}{N_{m+}! N_{m-}!}\left(\prod_a^{N_{m+} + N_{m-}} \int d^{d+1} x_a\right)\int [{\cal D}A^{\mbox{ \scriptsize ph}}_\mu]\exp\left[-\Gamma_{\mbox{\scriptsize eff}}\right]\,,
\label{partition function before compactifying}
\end{eqnarray}
and the effective action is given by
\begin{eqnarray}
\Gamma_{\mbox{\scriptsize eff}}=\mbox{Tr}\left[\log {\cal K}_{W^\pm} \right]+\mbox{Tr}\left[\log {\cal K}_{\phi^\pm} \right]-2\mbox{Tr}\left[\log {\cal K}_{c^\pm} \right]+\int d^{d+1}x\frac{1}{4g_3^2} F_{\mu\nu} F_{\mu\nu}\,,
\label{the effective action of at finite temperature}
\end{eqnarray}
where ${\cal K}_{W^\pm}=- D^2\delta_{\mu\nu}+{\cal M}_{\mu\nu}(x)$, ${\cal K}_{\phi^\pm}={\cal K}_{c^\pm}=- D^2+M_W^2$, and we neglected the determinant over the Higgs field $\phi$ since it can only modify the short range physics.
\footnote{The effect of a finite Higgs mass on the deconfinement transition was considered in \cite{Kovchegov:2002vi}.}
 The trace $\mbox{Tr}$ denotes the trace over both spacetime and Lorentz (Euclidean) indices. Thus we have $\mbox{Tr} \log {\cal K}=\mbox{tr}\int d^{d+1}x \left\langle x|\log{\cal K} |x\right\rangle$, where $\mbox{tr}$ denotes the trace over the Lorentz indices. At this stage, it is useful to use the zeta function regularization technique to express the trace $\log$s in terms of the diagonal matrix elements of the heat kernel $\left\langle x|e^{-\tau{\cal K}} |x\right\rangle$:
\begin{eqnarray}
\nonumber
&&-\mbox{Tr}\left[\log {\cal K}_{W^\pm} \right]-\mbox{Tr}\left[\log {\cal K}_{\phi^\pm} \right]+2\mbox{Tr}\left[\log {\cal K}_{c^\pm} \right]\\
\nonumber
&&=\frac{d}{ds}\left [\frac{1}{\Gamma(s)}\int _0^\infty d\tau\tau^{s-1} \int d^{d+1} x \mbox{tr}\left(\langle x| e^{-\tau {\cal K}_{W^\pm}}|x\rangle+\langle x|e^{-\tau {\cal K}_{\phi^\pm}}|x\rangle-2\langle x|e^{-\tau {\cal K}_{c^\pm}}|x\rangle\right)\right]_{s=0}\,.\\
\label{zeta function reg}
\end{eqnarray}

Now, we are in a position to calculate the effective action (\ref{partition function before compactifying}) upon compactifying one of the dimensions over a thermal circle $\S_\beta^1$. In fact, one just needs to calculate the heat kernels $\mbox{Tr}\left[e^{-\tau {\cal K}}\right]$ over the manifold $\R^d\times \S_\beta^1$. This will be achieved  in the next section.

\subsection{The heat kernel expansion}

The heat kernel expansion methods is a systematic way to carry out a series expansion of the operators  $\left\langle x|e^{-\tau{\cal K}} |x\right\rangle$ in powers of the parameter $\tau$ using only gauge invariant quantities as expansion coefficients. However, the $\tau$ expansion is local and hence is insensitive to the global structure of the background manifold. Therefore, one normally would expect the expansion coefficients to be  functions of the local gauge invariant quantity $F_{\mu\nu}$. Such an expansion can easily overlook non-trivial global gauge invariant structures on the background manifold. This is particularly clear in the case of gauge theories formulated on $\R^d\times \S^1_\beta$, where the manifold admits a gauge holonomy or the Polyakov loop $\Omega$ around the thermal $\S^1_\beta$ circle:
\begin{eqnarray}
\Omega=\exp\left [i\int_0^\beta dx_0  A_0(\vec x,x_0)\right]\,.
\label{the definition of Omega}
\end{eqnarray}
If we choose a gauge in which  $A_0$ is time independent, then we find $\Omega=e^{i\beta A_0}$. Thus, the Polyakov loop is basically the exponent of the zero frequency component of the gauge field $A_{0}$ along the thermal circle, which constitutes a global gauge invariant quantity. Therefore, one needs to carry out a systematic expansion in powers of $\tau$ with expansion coefficients as functions of the two gauge invariant quantities $F_{\mu\nu}$ and $\Omega$, keeping in mind that the Lorentz invariance of the theory is explicitly broken due to the presence of the heat path. Such a systematic approach was developed in \cite{Megias:2002vr,Megias:2003ui}.

Given a massless Klein-Gordon operator of the form ${\cal U}(x)-D_\mu^2$, where ${\cal U}(x)$ is a local function of $x$, the authors in \cite{Megias:2002vr,Megias:2003ui} showed that
\begin{eqnarray}
\mbox{Tr}\left[e^{-\tau \left({\cal U}(x)-\hat D^2\right)}\right]=\frac{1}{\left(4\pi \tau\right)^{(d+1)/2}}\sum_n\int d^{d+1}x \tr\left[b_n \right]\tau^n\,,
\label{heat kernel expansion}
\end{eqnarray}
where $\tr$ denotes the trace over the Lorentz indices.
\footnote{Also, see \cite{MoralGamez:2011en} for the heat kernel expansion for spacetimes with topology $\R^n\times \S^1\times...\times\S^1$. }
 Unlike  the case of non-compact spaces, where only integer powers of $\tau$ appear in the expansion, half-integer powers of  $\tau$ are also allowed in the present case. The coefficients $b_n$ are given, up to the second power of $\tau$, by
\begin{eqnarray}
b_0=\psi_0\,,b_{1/2}=0\,, b_1=-\psi_0{\cal U}\,, b_{3/2}=0\,, b_2=\psi_0\left(\frac{1}{2}{\cal U}^2+\frac{1}{12} F_{\mu\nu}^2 \right)-\frac{1}{6}(\psi_0+2\psi_2) E_i^2\,,
\label{bn coeff}
\end{eqnarray}
and the electric field $ E_i$ is defined as $ E_i= F_{0i}$. 
\footnote{Notice that the first non-zero half-integer coefficient is $b_{5/2}$ which we do not consider here; see \cite{Megias:2003ui}.}
The functions $\psi_n$ are periodic functions of the Polyakov loop 
\begin{eqnarray}
\psi_n=\frac{\sqrt{4\pi \tau}}{\beta}\sum_{\omega_n}\tau^{n/2}\left(i\omega_n -\frac{1}{\beta}\log \Omega \right)^n\exp\left[\left(i\omega_n -\frac{1}{\beta}\log \Omega \right)^2 \tau\right]\,,
\end{eqnarray}
where $\omega_n=2\pi n/\beta$ are the Matsubara frequencies. Notice that the potential along the $\S^1_\beta$ circle $A_0$ as well as the one-loop induced field strength $ F_{\mu\nu}$ have contributions from both the photon $A^{\mbox{\scriptsize ph}}_\mu$ and monopole $ {\cal A}_\mu$ fields, as was stressed above, i.e. $ A_0= A_0^{\mbox{\scriptsize ph}}+{\cal A}_0$, and   $ F_{\mu\nu}= F^{\mbox{\scriptsize ph}}_{\mu\nu}+{\cal F}_{\mu\nu}$\,.

Since our Klien-Gordon operators ${\cal K}_{W^\pm}$, ${\cal K}_{\phi^\pm}$, and ${\cal K}_{c^\pm}$ are massive, with mass $M_W$, we need to modify  the coefficients $b_n$ in (\ref{bn coeff}).  This can be done either by resumming the series in (\ref{heat kernel expansion}) or  slightly modifying the derivation in \cite{Megias:2003ui} that leads to (\ref{bn coeff}). Simply, noticing that ${\cal U}(x)=M_W^2+\bar {\cal U}(x)$, the $M_W^2$ part  in the coefficients $b_n$ is resummed to give $e^{-M_W^2\tau}$.  Then, the heat kernel expansion is given by the modified expression
\begin{eqnarray}
\mbox{Tr}\left[e^{-\tau \left(\bar{\cal U}(x)+M_W^2- D^2\right)}\right]=\frac{e^{-M_W^2\tau}}{\left(4\pi \tau\right)^{(d+1)/2}}\sum_n\int d^{d+1}x \tr\left[\bar b_n \right]\tau^n\,.
\label{ mod heat kernel expansion}
\end{eqnarray}
The coefficients $\bar b_n$ are given by (\ref{bn coeff}) after making the replacement ${\cal U} \rightarrow \bar{\cal U}$, and we have  $\bar{\cal U}=2 F_{\mu\nu}$ for ${\cal K}_{W^\pm}$ (the Lorentz structure is implicitly understood), and $\bar{\cal U}=0$ for both ${\cal K}_{\phi^{\pm}}$ and  ${\cal K}_{c^{\pm}}$.

Taking the trace over the Lorentz indices, we obtain up to the second power in $\tau$ in the heat kernel expansion
\begin{eqnarray}
\nonumber
\Gamma_{\mbox{\scriptsize eff}}&=&\int d^{d+1}x\left\{\frac{1}{4g_3^2} F_{\mu\nu} F_{\mu\nu} -\frac{d}{ds}\left[\frac{1}{\Gamma(s)} \int _0^\infty d\tau\int \frac{d^d p}{\left(2\pi\right)^d} \frac{e^{-\tau(M_W^2+p^2)}}{\sqrt{4\pi \tau}}\tau^{s-1}\right.\right.\\
&&\left.\left.\quad\quad \times \left[d\psi_0+\tau^2 \left(\psi_0\left(-2+\frac{d}{12}\right) F_{\mu\nu}^2-\frac{d}{6}(\psi_0+2\psi_2) E_i^2\right)  \right]\right]_{s=0}  \right\}\,,
\label{main effective action}
\end{eqnarray}
where we have used $\frac{1}{\left(4\pi \tau\right)^{d/2}}=\int \frac{d^d p}{\left(2\pi\right)^d}e^{-\tau p^2}$ which  proves to be useful in simplifying the sums over the Matsubara frequencies. These sums are evaluated in Appendix B. The expressions for the different terms are given by (\ref{psi0}), (\ref{tau2psi0}), and (\ref{tau2psi2}), where we also give the leading-order behavior in the limit $T<<M_W$.  Collecting everything and setting $d=2$, we find that the effective action in the limit $T<<M_W$ reads
\begin{eqnarray}
\nonumber
\Gamma_{\mbox{\scriptsize eff}}&=&\int d^{2+1}x \left\{\left(\frac{1}{4g_{3}^2}+\frac{11}{48\pi M_W}\right) F_{\mu\nu}^2\right.\\
\nonumber
&&\left.\quad\quad\quad+e^{-\frac{M_W}{T}}\left[-2\frac{T^2M_W}{\pi}+\frac{11}{24\pi M_W} F_{\mu\nu}^2+\frac{ E_i^2 }{12\pi T} \right]\cos\left(i\log\Omega\right)   \right\}\,.\\
\label{final expression for the effective action}
\end{eqnarray}
The term $11F_{\mu\nu}^2/(48\pi M_W)$ is a temperature-independent one-loop correction to the $U(1)$ free Lagrangian. Hence, we  defined the effective coupling constant as $1/g_{3\mbox{\scriptsize eff}}^2=1/g_3^2+11/(12\pi M_W)$. 

Now, a few remarks are in order:
\begin{enumerate}
\item The terms $11 F_{\mu\nu}^2/\left(24\pi M_W\right)$ and $E_i^2/\left(12\pi T\right)$ multiply the exponentially small factor $e^{-M_W/T}$. Therefore, both terms  can be neglected compared to the kinetic term $F_{\mu\nu}^2/\left(4g_{3\mbox{\scriptsize eff}}^2\right)$. This is unlike the term $e^{-M_W/T}T^2M_W\cos\left(i\log\Omega\right)/\pi$ which is responsible for highly non-trivial dynamics. As we show in the following section, the partition function (\ref{partition function before compactifying}) with the effective action
\begin{eqnarray}
\Gamma_{\mbox{\scriptsize eff}}=\int d^{2+1}x \left\{\frac{1}{4g_{3\mbox{\scriptsize eff}}^2} F_{\mu\nu}^2-\frac{2T^2M_W}{\pi}e^{-\frac{M_W}{T}}\cos\left(i\log\Omega\right)   \right\}
\label{reduced expression for the effective action}
\end{eqnarray}
represents a double Coulomb gas of electrically (W-bosons) and magnetically (monopole-instantons) charged particles. Each type of particles carry a positive or negative unit charge and interact via a long-range potential, while the electric and magnetic charges have the Aharonov-Bohm phase interaction. It is the $\cos\left(i\log\Omega\right)$ term in (\ref{reduced expression for the effective action}) that is responsible for the later type of interaction. 
\item It is important to emphasize  that the effective action (\ref{reduced expression for the effective action}) is a $U(1)$ gauge-invariant quantity.  This is obvious since both $F_{\mu\nu}$ and $\Omega$ are gauge-invariants. 
\item The effective action (\ref{reduced expression for the effective action}) description is valid for all temperatures in the range $0\leq T<M_W$.
\item The term  $2T^2M_We^{-M_W/T}\cos\left(i\log\Omega\right)/\pi$ is the leading order contribution coming from the full calculations of the effective action as presented in Appendix B. Higher order corrections will generally have the form $\sum_n{\cal P}_n(T/M_W)e^{-nM_W/T}\cos\left(in\log \Omega\right)$, where ${\cal P}_n(x)$ is a polynomial in $x$. Such terms describe complex molecules of W-bosons with higher mass and charge and in principle can be added to the Coulomb gas. However, such molecules have  highly suppressed dynamics thanks to the Boltzmann suppression factor $e^{-nM_W/T}$ for all molecules with $n>1$, and hence we ignore them in the following analysis.
\end{enumerate}

Next, we  proceed with our analysis to show that the partition function (\ref{partition function before compactifying}) with the effective action (\ref{reduced expression for the effective action}) represents a double Coulomb gas of electric and magnetic charges.

\subsection{The double Coulomb gas}

The first step in dealing with the partition function (\ref{partition function before compactifying}) is to manipulate the term   $\cos\left(i\log\Omega\right)$. Recalling that $A_0={\cal A}_0+A_0^{\mbox{\scriptsize ph}}$, as well as the definitions (\ref{total monopole field}) and (\ref{the definition of Omega}), and taking into account the fact that our theory is formulated on a thermal circle $\S_\beta^1$ (hence the monopole-instanton field is the result of the sum of an infinite number of images along the circle), we obtain
\begin{eqnarray}
\nonumber
\cos\left(i\log\Omega\right)&=&\cos\left(\int_0^\beta dx_0\sum_{a}q_a \sum_{n=-\infty}^\infty A_0^m(\vec x-\vec x_a, x_0-x_{0a}+n\beta )  +\int_0^\beta dx_0 A^{\mbox{\scriptsize ph}}_0\left(\vec x, x_0\right)\right)\\
&=&\cos\left(2\sum_a q_a \Theta\left(\vec x-\vec x_a\right)+\int_0^\beta dx_0A^{\mbox{ \scriptsize ph}}_0\left(\vec x, x_0\right)\right)  \,,
\end{eqnarray}
where we have used the trick in (\ref{manipulating the monopole field}), and the $\Theta$ angle is defined in (\ref{the angle theta}). Then, we use  $\cos \theta=\left(e^{i\theta}+e^{-i\theta}\right)/2$ to expand the terms $\exp\left(2\xi\int dx \cos\left( \theta(x)\right)\right)$ as follows
\begin{eqnarray}
\exp\left(2\xi\int dx \cos\left( \theta(x)\right)\right)=\sum_{n_+,n_-=0}^\infty\sum_{q_A=\pm 1}\frac{\xi^{n_++n_-}}{n_+!n_-!}\left(\prod_{A=1}^{n_++n_-} \int dx_A \right) e^{\sum_{A} iq_A \theta(x_A)}\,,
\label{the main identity for exponential}
\end{eqnarray}
where $q_A=\pm 1$. We apply the expansion (\ref{the main identity for exponential}) to the term $\cos\left(i\log\Omega\right)$ to find that the partition function (\ref{partition function before compactifying}) reads
\begin{eqnarray}
\nonumber
{\cal Z}_{\mbox{\scriptsize grand}}&=&\sum_{N_{m\pm}, q_a=\pm 1 }\sum_{N_{W\pm}, q_A=\pm 1 }\frac{\xi_m^{N_{m+} + N_{m-}}}{N_{m+}! N_{m-}!}\frac{(T\xi_W)^{N_{W+} + N_{W-}}}{N_{W+}! N_{W-}!}\\
\nonumber
&&\times \left( \prod_a^{N_{m+} + N_{m-}} \int d^{2+1} x_a\right) \left(\prod_A^{N_{W+} + N_{W-}} \int d^{2+1}x_A \right) \exp\left[2i\sum_{aA}q_aq_A\Theta\left(\vec x_a-\vec x_A\right)\right]\\
\nonumber
&&\times \int \left[{\cal D} A_\mu^{\mbox{\scriptsize ph}}\right] \exp\left\{- \int d^{2+1}x\left[ \frac{1}{4g_{3\mbox{\scriptsize eff}}^2}\left( F^{\mbox{\scriptsize ph}}_{\mu\nu}+{\cal F}_{\mu\nu}\right)^2-i\sum_A q_A A_0^{\mbox{\scriptsize ph}}(\vec x, x_0)\delta(\vec x-\vec x_A) \right]\right\}\,,\\
\label{semi final expression for Z}
\end{eqnarray}
where $\xi_W=\frac{TM_W}{\pi}e^{-M_W/T}$ is the W-boson fugacity. 
\footnote{This fugacity can also be obtained by integrating the Boltzmann distribution of a single W-boson $e^{-H/T}$, where $H=M_W+\frac{p^2}{2M_W}$, over the particle momenta:
\begin{eqnarray}
\xi_W=S_W\int \frac{d^2 p}{\left(2\pi\right)^2}e^{-\frac{M_W}{T}-\frac{p^2}{2M_WT}}=S_W\frac{TM_W}{2\pi}e^{-\frac{M_W}{T}}\,,
\end{eqnarray}
where $S_W=2$ is the spin degeneracy factor of the W-bosons.
}
Since the expansion in  $\vec x_A$ goes hand in hand with the expansion in the W-boson fugacity $\xi_W$, it is natural to interpret $\vec x_A$ as the positions of these W-bosons and $q_A$ as their charges. In fact, as we show below, it  turns out that this is the correct interpretation of $\vec x_A$ and $q_A$.

Next, we turn to the calculations of the path integral over $A^{\mbox{\scriptsize ph}}_\mu$.  This path integral is plagued by the presence of monopole-instantons. The reason is that the potential  $A^{\mbox{\scriptsize ph}}_\mu$, which naturally describes the force mediation between W-bosons, should also describe the force between monopole-instantons. However, as we mentioned in section 2, such a description is only possible on the expense of allowing Dirac-like singularities to appear in the gauge potential which can lead to inconsistencies. In other words, one has to exercise caution while carrying out the path integral over the field $A^{\mbox{\scriptsize ph}}_\mu$ which, in our case, is used to describe both electric and magnetic forces. In the absence of magnetic objects, the $U(1)$ gauge potential $A^{\mbox{\scriptsize ph}}_\mu$ is a genuine tool to describe the electric force between  electrically charged particles. We call this $U(1)$  the  abelian electric group. On the other hand, in the presence of magnetic charges, and absence of any electric objects,  one needs to go to a dual description to avoid the Dirac singularities that appear when using the electric $U(1)$. The  dual fields are invariant under a dual $U(1)$ that is known as the  abelian magnetic group.  The dual description  can be performed by  introducing a Lagrange multiplier field $\sigma$ as $\delta S=\frac{1}{2}\int d^3 x \sigma \epsilon_{\mu\nu\lambda}\partial_\mu F_{\nu\lambda}$ to the action $S=\int d^3 x\frac{1}{4g_3^2}F_{\mu\nu}^2$  to enforce the Bianchi identity $\epsilon_{\mu\nu\lambda}\partial_\mu F_{\nu\lambda}=0$ everywhere except  at the position of the magnetic charges.  Then, we vary $S+\delta S$ with respect to $F_{\mu\nu}$ to find $F_{\mu\nu}=-g_{3}^2\epsilon_{\mu\nu\lambda}\partial \sigma_\lambda$  and substitute back into  $S+\delta S$. The result is the action $S+\delta S=\int d^3 x \frac{g_{3}^2}{2}\left(\partial_\mu \sigma\right)^2$ in terms of the dual photon field $\sigma$. This dual description was used in the original work of Polyakov \cite{Polyakov:1976fu} to account for the monopole-instanton plasma.

 Fortunately enough, there exists a duality transformation prescription that enables us to perform the integral over $A^{\mbox{\scriptsize ph}}_\mu$ without having to run into contradictions even in the presence of both electric and magnetic charges \cite{Deligne:1999qp}. In this prescription, one enlarges the electric $U(1)$ gauge symmetry to $U(1)\times U(1)$. It is this extra gauge redundancy that enables us to perform the path integral over both electric and magnetic objects without running into difficulties. 
\footnote{I am grateful to Erich Poppitz for elucidating this method. }
 As a warm up exercise, we first show that the path integral $\int \left[{\cal D} A_\mu\right]\exp\left[ -\int d^3 x\frac{1}{4g_3^2} F_{\mu\nu}^2\right]$ is equivalent to the path integral  $\int \left[{\cal D} A_\mu\right]\left[{\cal D} \Phi_\nu\right]\left[{\cal D} \vartheta\right]\exp\left[ -\int d^3 x {\cal L}_{U(1)\times U(1)}\right]$ where 
\begin{eqnarray}
{\cal L}_{U(1)\times U(1)}=\frac{g_3^2}{2}\left(\partial_\mu \vartheta+\Phi_\mu\right)^2+i\epsilon_{\mu\nu\lambda}\partial_\mu A_\nu \Phi_\lambda\,.
\label{the double U Lagrangian}
\end{eqnarray} 
This Lagrangian is invariant under two different $U(1)$s: the first $U(1)$ is given by $\Phi_\mu \rightarrow \Phi_\mu+\partial_\mu \lambda$ and $\vartheta \rightarrow \vartheta-\lambda$, while the second $U(1)$ is $A_\mu \rightarrow A_\mu+\partial_\mu \sigma$. Setting $\vartheta=0$ (or in other words, using the unitary gauage), and varying ${\cal L}_{U(1)\times U(1)}$ with respect to $\Phi_\mu$ we find that the equation of motion of $\Phi_\mu$ reads $\Phi_\mu=-i\epsilon_{\mu\nu\alpha}\partial_\nu A_\alpha/g_3^2$. Then, substituting $\Phi_\mu$ back into ${\cal L}_{U(1)\times U(1)}$ we obtain ${\cal L}_{U(1)\times U(1)}=\frac{1}{4g_3^2}F_{\mu\nu}^2$ which proves the equivalence of the above  mentioned path integrals  to each other.
 
Now, we come to the path integral over $A_\mu^{\mbox{\scriptsize ph}}$ in (\ref{semi final expression for Z}).  According to the duality transformation prescription, we trade the Lagrangian (\ref{the double U Lagrangian})  for the term $F_{\mu\nu}^2/4g_{3\mbox{\scriptsize eff}}^2$ which results from expanding the square  in the last line in (\ref{semi final expression for Z}). Thus, we express the path integral over $A_{\mu}^{\mbox{\scriptsize ph}}$ as a double path integral over $A_{\mu}^{\mbox{\scriptsize ph}}$ and the additional auxiliary field $\Phi_\nu$:
\begin{eqnarray}
\int \left[{\cal D} A_\mu^{\mbox{\scriptsize ph}}\right]\left[{\cal D}\Phi_\nu \right] e^{-S_{\mbox{\scriptsize aux}}}\,,
\end{eqnarray}
where
\begin{eqnarray}
\nonumber
S_{\mbox{\scriptsize aux}}&=&\int d^{2+1}x\frac{1}{4g_{3\mbox{\scriptsize eff}}^2}\left({\cal F}_{\mu\nu}\right)^2+\frac{1}{2g_{3\mbox{\scriptsize eff}}^2}{\cal F_{\mu\nu}}F^{\mbox{\scriptsize ph}}_{\mu\nu}+ \frac{g_{3\mbox{\scriptsize eff}}^2}{2}\Phi_\mu^2\\
&&+i\epsilon_{\mu\nu\lambda}\partial_\mu A^{\mbox{\scriptsize ph}}_\nu \Phi_\lambda-i\sum_A q_AA_0^{\mbox{\scriptsize ph}}(\vec x,x_0)\delta(\vec x-\vec R_A)\,. 
\label{auxiliary Lagrangian}
\end{eqnarray}
Varying $S_{\mbox{\scriptsize aux}}$ with respect to $A_\mu^{\mbox{\scriptsize ph}}$ and substituting the resulting equation of motion back into (\ref {auxiliary Lagrangian}), we obtain the final expression of the grand partition function (the details of the procedure are presented in Appendix C): 
\begin{eqnarray}
\nonumber
{\cal Z}_{\mbox{\scriptsize grand}}&=&\sum_{N_{m\pm}, q_a=\pm 1 }\sum_{N_{W\pm}, q_A=\pm 1 }\frac{\xi_m^{N_{m+} + N_{m-}}}{N_{m+}! N_{m-}!}\frac{(T\xi_W)^{N_{W+} + N_{W-}}}{N_{W+}! N_{W-}!}\prod_a^{N_{m+} + N_{m-}} \int d^{2+1}x_a\prod_A^{N_{W+} + N_{W-}} \int \frac{d^2x_A}{T}\\
\nonumber
&&\times \exp \left[-\frac{8\pi^2}{g_{3\mbox{\scriptsize eff}}^2}\sum_{a,b}q_aq_bG(x_a- x_b)+\frac{g_{3\mbox{\scriptsize eff}}^2}{4\pi T}\sum_{A,B}q_Aq_B\log T|\vec x_A-\vec x_B|+2i\sum_{aA}q_aq_A\Theta\left(\vec x_a-\vec x_A\right)\right]\,.\\
\label{final expression for Z}
\end{eqnarray}
The Green's function $G$ satisfies $\nabla^2 G(x-x')=-\delta^3(x-x')$, where the Laplacian $\nabla^2$ is defined over $\R^2\times \S^1_\beta$. The Green's function is given by
\begin{eqnarray}
G(x- x')=\left(\frac{1}{4\pi|x-x'|}\right)^{(p)}=\frac{1}{4\pi}\sum_{n=-\infty}^{\infty}\frac{1}{\sqrt{\left(\vec x-\vec x'\right)^2+(x_0-x_0'+n\beta)^2}}\,.
\label{The Greens function interaction}
\end{eqnarray}

The partition function (\ref{final expression for Z}) describes a grand-canonical distribution of a three-dimensional double Coulomb gas of W-bosons and monopole-instantons at finite temperature $0 \leq T <M_W$. The W-bosons are physical particles and therefore they sweep world-lines as time elapses. These bosons interact logarithmically at all temperatures which is the expected behavior for particles in $2+1$ D. On the contrary, the monopole-instantons are pseudo-particles; they represent localized events which interact via (\ref{The Greens function interaction}). In addition, the W-bosons and monopole-instantons interact via the Aharonov-Bohm phase $\Theta(\vec x_a-\vec x_A)$. A picture of this gas is depicted in Figure (\ref{the 3 dimensional Coulomb gas picture}). 

 \begin{FIGURE}[ht]
    {
    \parbox[c]{\textwidth}
        {
        \begin{center}
        \includegraphics[angle=0, scale=.8]{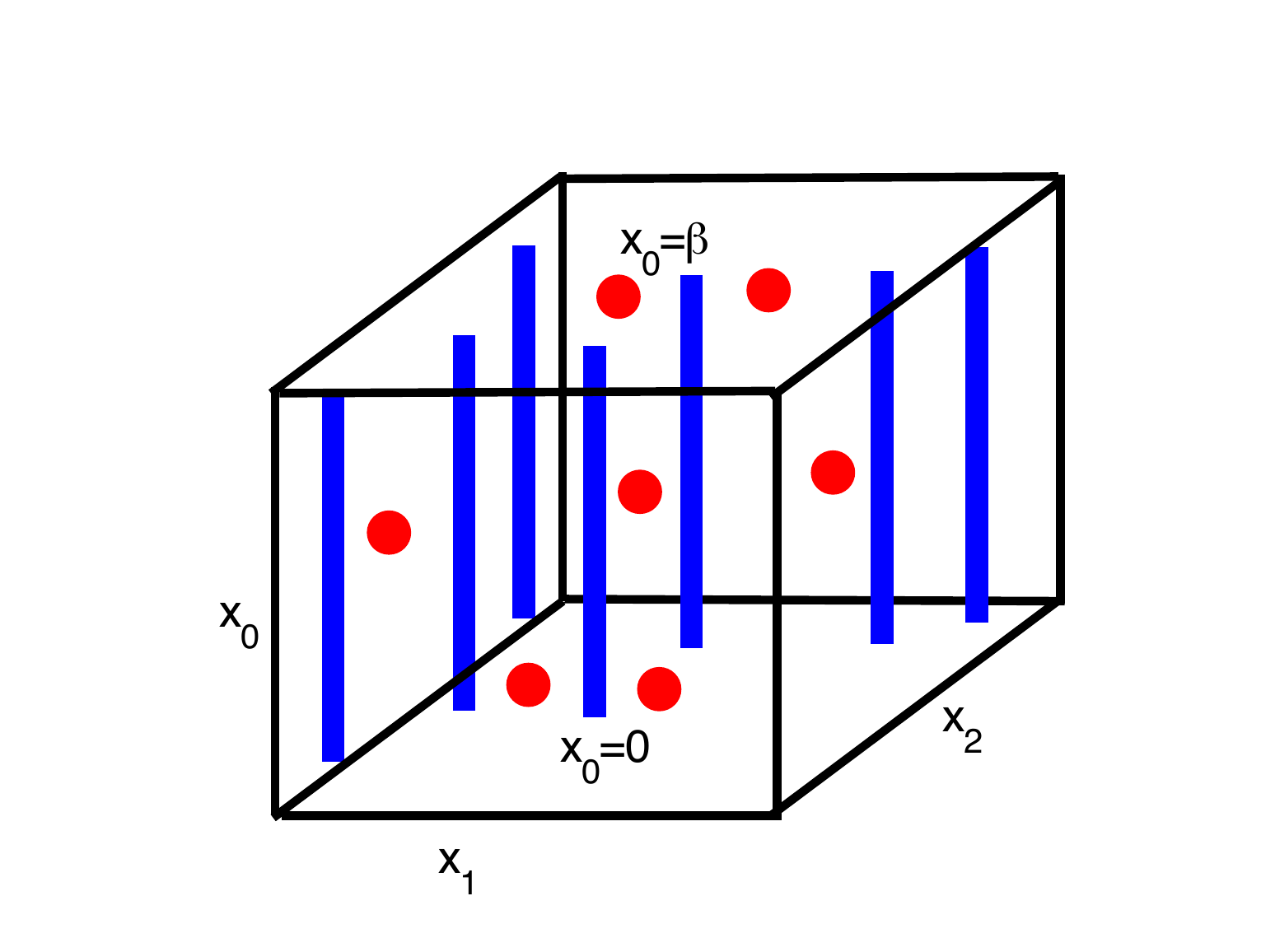}
	\hfil
        \caption 
      {A picture of the three-dimensional double Coulomb gas of the partition function (\ref{final expression for Z}). The W-bosons are vortices that extend along the $x_0$ direction, while the monopole-instantons are point-like particles. The planes $x_0=0$ and $x_0=\beta$, where $\beta$ is the inverse temperature, are identified.
			}
      \label{the 3 dimensional Coulomb gas picture}
       \end{center}
        }
    }
\end{FIGURE}

\subsection{Obtaining the Coulomb gas from the non-relativistic partition function}

The  non-relativistic partition function (\ref{non relativistic total partition function}), along with the Lagrangian (\ref{total non relativistic Lagrangian}), is readily in the form of a Coulomb gas. Comparing (\ref{non relativistic total partition function}) with (\ref{final expression for Z}), we see the expected form of interaction between the same species: the logarithmic potential between the W-bosons as well as the $G(x-x')$ interaction between the monopole-instantons.  What remains to be shown is the Aharonov-Bohm phase interaction between the W-bosons and monopole-instantons. To show that such term arises naturally from (\ref{non relativistic total partition function}) we proceed as follows. First, we notice that the term $\int d^2 x' \int_0^\beta dx_0' \vec A^{m(p)}(\vec x-\vec x', x_0-x_0')\rho_m(\vec x', x_0')$, which accompanies the kinetic factor $\vec \nabla$,  is suppressed by the W-boson mass $M_W$ compared to the term $-ieg_m\int d^2 x\int d^2x'\int_0^\beta dx_0'\left[\rho_{W_+}(\vec x,x_0)-\rho_{W_-}(\vec x,x_0) \right]A_0^{m(p)}\left(\vec x-\vec x',x_0-x_0'\right)\rho_m(\vec x',x_0')$, and hence we ignore it. Also, we leave aside the Coulomb interaction terms in (\ref{total non relativistic Lagrangian}), which are left intact in our present treatment.  Then, we have for the remaining part of the partition function
\begin{eqnarray}
Z_{\mbox{\scriptsize phase}}&=&\int \left[D\Phi_+\right]_\beta\left[D\Phi_-\right]_\beta\left[D\Phi^*_+\right]_\beta\left[D\Phi_-^*\right]_\beta \exp\left[-S_{\mbox{\scriptsize phase}}\right]\,,
\label{the phase partition function}
\end{eqnarray}
where
\begin{eqnarray}
\nonumber
S_{\mbox{\scriptsize phase}}&=&\int_0^\beta dx_0\int d^2x\Phi_+^*\left[M_W+\partial_{x_0}-\frac{\nabla^2}{2M_W} \right]\Phi_++\int_0^\beta dx_0 \int d^2 x\Phi_-^*\left[M_W+\partial_{x_0}-\frac{\nabla^2}{2M_W} \right]\Phi_-\\
\nonumber
&& +ieg_m\int d^2 x\int d^2x'\int_0^\beta dx_0\int_0^\beta dx_0'\left[\rho_{W_+}(\vec x,x_0)-\rho_{W_-}(\vec x,x_0) \right]A_0^{m(p)}\left(\vec x-\vec x',x_0-x_0'\right)\rho_m(\vec x',x_0')\,.\\
\end{eqnarray}
Using (\ref{definition of rhoW}) and   (\ref{monopole density operator}) to express the operators $rho_{W_\pm}$ and $\rho_m$ in terms of the defining Dirac-delta functions, and using (\ref{monopole field}) to perform the integral over $x_0'$, noticing that $eg_m=1$, we find
\begin{eqnarray}
\nonumber
S_{\mbox{\scriptsize phase}}&=&\int_0^\beta dx_0\int d^2x\Phi_+^*\left[M_W+\partial_{x_0}-\frac{\nabla^2}{2M_W} +2i\sum_a q_a\Theta(\vec x-\vec x_a)\right]\Phi_+\\
&+&\int_0^\beta dx_0\int d^2x\Phi_-^*\left[M_W+\partial_{x_0}-\frac{\nabla^2}{2M_W} -2i\sum_a q_a\Theta(\vec x-\vec x_a)\right]\Phi_-\,.
\label{final expression for the phase action}
\end{eqnarray}
Since the action (\ref{final expression for the phase action}) is quadratic in the fields $\Phi_+$ and $\Phi_-$, we can perform the path integral exactly to obtain
\begin{eqnarray}
\nonumber
Z_{\mbox{\scriptsize phase}}&=&\mbox{Det}\left[M_W+\partial_{x_0}-\frac{\nabla^2}{2M_W} +2i\sum_a q_a\Theta(\vec x-\vec x_a) \right]^{-S_W}\\
&&\times \mbox{Det}\left[M_W+\partial_{x_0}-\frac{\nabla^2}{2M_W} -2i\sum_a q_a\Theta(\vec x-\vec x_a) \right]^{-S_W}\,,
\end{eqnarray}
where $S_W$ is the spin degeneracy factor of the W-boson which is $2$ in $2+1$ D. Therefore, the partition function (\ref{the phase partition function}) can be casted in the form $Z_{\mbox{\scriptsize phase}}=e^{-\Gamma_{\mbox{\scriptsize phase eff}}}$, where
\begin{eqnarray}
\nonumber
\Gamma_{\mbox{\scriptsize phase eff}}=S_W\sum_{n=-\infty}^{\infty} T\int d^2x\int \frac{d^2k}{\left(2\pi\right)^2}\log\left[\left(\omega_n+2\sum_{a}q_a \Theta(\vec x-\vec x_a)\right)^2+\left(M_W+\frac{k^2}{2M_W}\right)^2 \right]\,.\\
\label{the sum and integral phase eff}
\end{eqnarray}
The expression of $\Gamma_{\mbox{\scriptsize phase eff}}$ is in the form of  the non-relativistic limit of  (\ref{the first integral in appendix B}). Following the same procedure that led from (\ref{the first integral in appendix B})  to (\ref{psi0}), we find 
\begin{eqnarray}
\Gamma_{\mbox{\scriptsize phase eff}}=-\frac{2M_W T}{\pi}\int d^2 x e^{-M_W/T}\cos\left(2\sum_a q_a \Theta(\vec x-\vec x_a)\right)\,.
\end{eqnarray}
Finally, using  (\ref{the main identity for exponential}) and restoring the Coulomb interaction terms, we recover the partition function of the double Coulomb gas (\ref{final expression for Z}), apart from the trivial renormalization $g_3 \rightarrow g_{3\mbox{\scriptsize eff}}$. In fact, these calculations are a non-trivial check on our many-body formulation presented in Section 3.

\subsection{The two-dimensional Coulomb gas and deconfinement transition}

As we mentioned above, the partition function (\ref{final expression for Z}) encodes all information about  the double Coulomb gas for  temperatures in the range $0 \leq T < M_W$.  At zero temperature, the monopole-insatntons proliferate and form a plasma of magnetic charges. This works as a dual Meissner effect which screens the magnetic field lines. The monopole-instanton fugacity is $\xi_m \sim e^{-4\pi v/g_3}$, where here and in the following analysis we omit dimensionfull pre-exponential factors and work in the BPS limit.  Thus, the average distance between two monopoles in the plasma is $d_{m-m}\sim e^{2\pi v/(3g_3)}$. Then, two external electric probes separated a distance $R>> d_{m-m}$ experience a linear confining potential $V=\sigma R$, where $\sigma\sim e^{2\pi v/(g_3)}$ is the string tension.  This continues to be the case as we slightly increase the temperature of the gauge theory. However, at any non-zero temperature, the W-bosons will start to proliferate according to the Boltzmann distribution $e^{-M_W/T}$. 

At low temperatures,  the W-bosons will form  electrically neutral dipoles which  screen the electric flux lines of the external probes resulting in a decrease of the string tension. The average size of the $W^+$-$W^-$ dipoles can be determined  using simple classical statistical arguments. First, we model the potential between the $W^+$ and $W^-$ constituents of the dipole by
\begin{eqnarray}
V(r)=\frac{g_3^2}{2\pi }\log(Tr) +\sigma r\,.
\end{eqnarray}
A better model would use the potential (\ref{The potential in terms of Meijer G function}). However, this adds  unnecessary complications to our problem. The average size of the $W^+$-$W^-$ molecule  is given by
\begin{eqnarray}
\langle r\rangle_{W_-W_+}=\frac{\int d^2x re^{-V(r)/T}}{\int d^2x e^{-V(r)/T}}=\frac{\int_{1/L}^\infty dr r^{2-2T_c/T}e^{-\sigma r/T}}{\int_{1/L}^\infty dr r^{1-2T_c/T}e^{-\sigma r/T}}\,,
\label{statistical mechanical expression for rW}
\end{eqnarray}
where $T_c=g_3^2/(4\pi)$ is the deconfinement temperature, as we argue below, and $L$ is a UV cutoff. At low temperatures, $0<T<<T_c$, we expand $r_{W_-W_+}$ about $T=0$ to find
\begin{eqnarray}
\langle r \rangle_{W_-W_+}|_{T \rightarrow 0}=L+\frac{T}{\sigma}\,.
\end{eqnarray}
This means that at low temperatures, the average size of the W molecules is determined by the UV cutoff of the theory, $L$, where the logarithmic interaction dominates over the linear confinement potential. At temperatures $T\sim T_c$, the average size of the $W_+-W_-$ molecules can be obtained by expanding (\ref{statistical mechanical expression for rW}) about $T=T_c$ 
\begin{eqnarray}
\langle r\rangle_{W_-W_+}|_{T \rightarrow T_c}\sim  \frac{T}{\sigma}e^{-\frac{L\sigma}{T}} \sim TL^2 e^{\frac{2\pi v}{g_3}}\,.
\end{eqnarray}
Hence, at temperatures $\sim T_c$, the linear potential dominates over the logarithmic interaction and we can visualize the gas as neutral molecules of W-bosons bound due to the existence of long flux tubes or in other words strings. This behavior is illustrated in Figure \ref{molecule size}.
  \begin{FIGURE}[ht]
    {
    \parbox[c]{\textwidth}
        {
        \begin{center}
        \includegraphics[angle=0, scale=1]{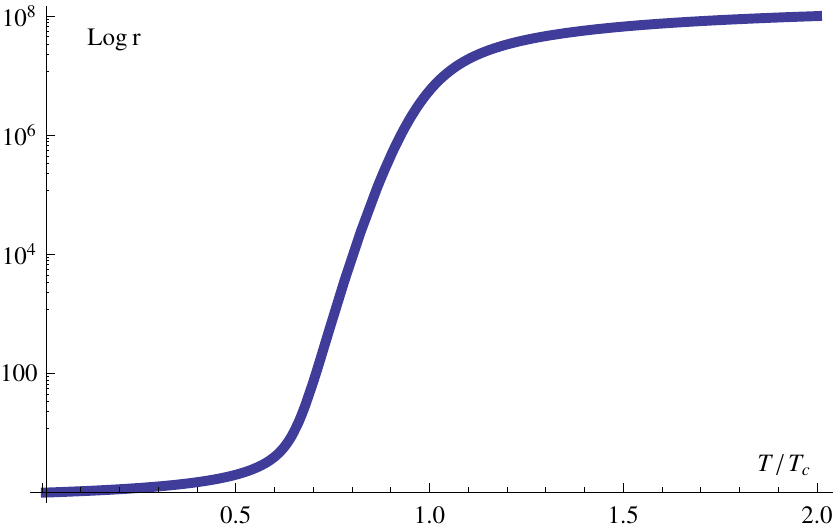}
	\hfil
        \caption 
      {  Behavior of the radius of $W_+W_-$ molecule as a function of temperature. We used $\sigma \sim 10^{-8}$ for the purpose of illustrating the transition between scales. We see that at $T<<T_c$ the average size is of the same order of the UV cutoff, while fatty molecules start to appear just below $T_c$. The right part of the diagram above $T_c$ is not trusted since the mass gap vanishes and the W-bosons deconfine. 
			}
        \label{molecule size}
        \end{center}
        }
    }
\end{FIGURE}
 On the other hand, the number density of the W-bosons is $n_W\sim e^{-\frac{M_W}{T}}\sim e^{-4\pi v T_c/(g_3 T)}$. Hence, the average distance between two W-bosons is $d_W\sim e^{2\pi vT_c/(g_3T)}$. Thus, we have $\langle r \rangle_{W_-W_+}\leq d_W$ for all $T\leq T_c$ which justifies the above picture of having a dilute gas of W-molecules. Near the transition temperature from below $T\cong T_c$, the separation between the  W-bosons inside the molecules $\langle r \rangle_{W_-W_+}$ becomes comparable to  $d_{W}$, and therefore we basically have a gas of free W-bosons. This justifies that $T_c$ is the critical deconfinement temperature. 

Such picture can be confirmed more rigorously  by studying the phase transition of the double Coulomb gas in (\ref{final expression for Z}). At temperatures much higher than the inverse distance between monopole-instantons, $T>>L^{-1}e^{-2\pi v/(3g_3)}$, yet much lower than the deconfinement transition temperature, $T_c$, the effective interaction between two monopoles is two-dimensional, and one has for the Green's function that appears in (\ref{final expression for Z}) $G(x-x')\rightarrow -T\log|\vec x-\vec x'|/(2\pi)$. Thus, the double Coulomb gas (\ref{final expression for Z})  becomes essentially two-dimensional. Such  a two-dimensional double Coulomb gas was studied previously in \cite{Dunne:2000vp}. This was done by mapping the gas to a dual-sine-Gordon model. Indeed, it was found that a second-order phase transition takes place at $T_c$. We do not elaborate further on the nature of the transition, and we refer the reader to the original literature for more details.

\section{Conclusion and future directions}

In this work, we have studied several issues of the confinement problem in the Georgi-Glashow model in $2+1$ D. The aim of our study was two-fold. First, we worked out a many-body description of the effective degrees of freedom, namely W-bosons and monopole-instantons,  for temperatures in the range $0\leq T<M_W$, where the W-bosons are non-relativistic. In this approach, we wrote down a partition function for the W-bosons in the background of an arbitrary number of monopole-instantons. This partition function, with the aid of a mean-field approximation, enabled us to find an explicit expression for the potential between two external electric probes at all distances. 

Then, we used a systematic method to integrate out the W-bosons at finite temperatures. We started with the relativistic partition function  in the background of the instantons field and applied a heat kernel expansion technique that takes into account the existence of a non-trivial thermal holonomy. We found that the partition function describes a three-dimensional two-component Coulomb gas; these are W-bosons and Monopole-instantons. The W-bosons interact logarithmically, while the monopoles interact via the potential $G(x)$, where $G(x)$ is the Green's function of the Laplacian operator on $\R^2\times \S^1_\beta$ and $S^1_\beta$ is the thermal circle. In addition, there is the Aharonov-Bohm phase interaction between the W-bosons and monopole-instantons. Further, we used the finite temperature many-body partition function to arrive to the same picture, which works as an independent check on our many-body methods.  For temperatures much larger than the inverse distance between two neighboring monopoles, the Coulomb gas becomes two-dimensional and we recover the previous result of \cite{Dunne:2000vp}.

The Georgi-Glashow model  has  been an important testing ground for the problem of confinement. In the following, we consider a few venues where our methods can be applied:
\begin{enumerate}
\item Our approach, of starting from the relativistic partition function and then integrating out the heavy fields, can be extended to include additional fields as well. In principle, one can consider fundamental scalars \cite{Dunne:2002rt} or fermions \cite{Antonov:2003rj}, or one can add a chemical potential to the system.
\item  Our many-body approach can be used to study the non-equilibrium phenomena in the Gerorgi-Glashow model, see e.g. \cite{nonequilibrium}. This can shed light on the nature of Yang-Mills in the non-equilibrium state. Recently, a tremendous effort has been made to calculate various kinetic coefficients, e.g. shear viscosity, which is important to understand the quark-gluon state of matter.  
\item One way of understanding the the color confinement mechanism in $4$-D pure Yang-Mills theory is through the use of abelian monopoles, see e.g. \cite{Diakonov:2007nv,Bruckmann:2011yd}. Moreover, it has been shown that such monopoles can play a prominent role in understanding the deconfinement phase transition in this theory \cite{D'Alessandro:2010xg}. It will be interesting to use a many-body approach to tackle this problem along the same lines presented in the present work.
\item An important class of theories is $SU(N)$  Yang-Mills on $\R^{2,1}\times \S^1$, where $\S^1$ is a spatial circle. These theories abelianize either by adding deformations \cite{Unsal:2008ch}, or by adding adjoint fermions with periodic boundary conditions along the spatial  circle \cite{Unsal:2007jx}.  Integrating out the heavy Kaluza-Klein tower results in an effective potential for the gauge-component along the $\S^1$ direction, which can be thought of as a compact adjoint scalar. This reduces the $SU(N)$  Yang-Mills to  a Georgi-Glashow model, possibly with adjoint fermions. In \cite{Anber:2011gn}, it was shown that $SU(N)$ Yang-Mills with adjoint fermions on $\R^{2,1}\times \S^1$, near the confinement-deconfinement transition region, can be mapped to two-dimensional "affine" XY spin models. These spin models can be studied analytically or numerically, by means of Monte Carlo simulations \cite{Anber:2012ig}, which can shed light on the nature of the deconfinement transition in Yang-Mills in $3+1$ D. The mapping between the gauge theories and spin models was possible by showing that the partition function of both systems is that of a two-dimensional two-component Coulomb gas. Regarding (1) above, our methods provide a systematic way of deriving the Coulomb gas partition function even in the presence of an additional scalar or fermion fields, which can arise in this class of theories.  
\end{enumerate}

\section*{Acknowledgment}

I would like to thank Erich Poppitz for collaboration at the early stages of this project, for many enlightening discussions and comments on the manuscript. I would also like to thank Mithat {\"U}nsal for valuable comments on the manuscript. I am also thankful to Yulia Smirnova for editing the manuscript.  This work has been supported by NSERC Discovery Grant of Canada.

\appendix

\section{The Wilson loop calculations}

In this section, we derive the result (\ref{the string tension using the mean field approximation}) by performing a perturbative treatment to the expectation value of the Wilson loop. Before performing these calculations, we first review the steps that lead to the  exact result of the string tension that was obtained by Polyakov \cite{Polyakov:1976fu}. Then, we redo the Polyakov's calculations using a Gaussian approximation. This introduces an error in the value of the string tension due to the neglection of the non-Gaussianities. This value of the string tension coincides with the value we obtained in Section 3 using the many-body approach.  

\subsection{The exact Polyakov calculations}

The exact partition function of the monopole-instanton gas was obtained by Polyakov in a seminal paper \cite{Polyakov:1976fu}:
\begin{eqnarray}
\label{Polyakov action}
{\cal Z}_{\mbox{\scriptsize sin-Gordon}}=\int \left[D\sigma\right] \exp\left[-\frac{1}{2}\left(\frac{g_3}{4\pi}\right)^2\int d^3 x \left[\left(\nabla \sigma\right)^2-2{\cal M}^2\cos\sigma \right]\right]\,,
\label{the main equation for polyakov treatment}
\end{eqnarray}
where $\sigma$ is the dual photon field, and  ${\cal M}$ is its mass. In order to find the string tension, we calculate the expectation value of the Wilson loop:
\begin{eqnarray}
\left \langle W(C) \right \rangle= \left\langle \exp\left[i \oint_C A_\mu^m dx^\mu\right] \right\rangle = \left\langle \exp\left[i \int _S B_\mu^m dS^\mu\right] \right\rangle \,,
\end{eqnarray}
where $C$ and $S$ are respectively a closed loop and the surface enclosed by it, and the equality in the above equation is a result of using the Gauss's theorem.  The magnetic field $B_\mu^m$ is sourced by the monopole-instantons:
\begin{eqnarray}
B^m_\mu( x) =\epsilon_{\mu\alpha\beta}\partial_\alpha A_\beta^m=\int d^3y \frac{( x- y)_\mu\rho ( y)}{| x- y|^3}\,.
\end{eqnarray}
Hence, the Wilson loop can be written as
\begin{eqnarray}
\label{ the ultimate integral}
\left \langle W(C) \right \rangle= \left\langle \exp\left[i \int d^3 x \eta ( x) \rho( x) \right] \right\rangle \,,
\end{eqnarray}
where
\begin{eqnarray}
\label{eta integral}
\eta(x)=\int_{S_y} d S_y \cdot\frac{( x-y)}{|x-y|^3}\,.
\end{eqnarray}
The quantity $\eta(x)$ is the magnetic flux through the surface $S$ due to a unit magnetic charge located at position $x$ in the Euclidean spacetime.  Including the Wilson loop into (\ref{the main equation for polyakov treatment}), one finds
\footnote{We refer the reader to the original literature \cite{Polyakov:1976fu} for details.} 
\begin{eqnarray}
\label{Polyakov action 2}
\left \langle W(C) \right \rangle=\exp\left[-\frac{1}{2}\left(\frac{g_3}{4\pi}\right)^2\int d^3 x \left[\left(\nabla (\sigma-\eta)\right)^2-2{\cal M}^2\cos\sigma \right]\right]\,.
\end{eqnarray}
The dominant contribution to $\left \langle W(C) \right \rangle$ comes from the classical solution to the effective action. Thus, one extremizes $\left \langle W(C) \right \rangle$ to obtain the sine-Gordon equation
\begin{eqnarray}
\nabla^2\left(\sigma-\eta\right)={\cal M}^2\sin\sigma\,.
\end{eqnarray}
We can choose the loop to be an infinite loop that lies on the $x_0-x_2$ plane. Hence, the above equation reduces to
\begin{eqnarray}
\partial^2_{x_1}\sigma=4\pi \delta'(x_1)+{\cal M}^2\sin \sigma\,,
\label{domain wall equation}
\end{eqnarray}
where we have used  $\partial^2_{x_1} \eta=4\pi \delta'(x_1)$. Solving for $\sigma$ and substituting back  into $\left \langle W(C) \right \rangle$, one obtains the string tension
\begin{eqnarray}
\sigma_{\mbox{\scriptsize string-exact-Polyakov}}=\frac{g_3^2}{2\pi^2}{\cal M}\,.
\label{exact string tension from Polyakov calculations}
\end{eqnarray}
%

\subsection{The Guassian approximation to the Polyakov model}

In this section, we redo the Polyakov calculations using the Gaussian approximation (or mean-field approximation), i.e. taking $\cos\sigma \cong1 -\frac{1}{2}\sigma^2$. In this case, the expectation value of the Wilson loop is
\begin{eqnarray}
\label{Polyakov action 2 Gaussian}
\left \langle W(C) \right \rangle=\exp\left[-\frac{1}{2}\left(\frac{g_3}{4\pi}\right)^2\int d^3 x \left[\left(\nabla (\sigma-\eta)\right)^2+{\cal M}^2\sigma^2 \right]\right]\,.
\label{the Wilson line in the Gaussian approximation}
\end{eqnarray}
 The corresponding  domain-wall equation (\ref{domain wall equation}) reduces to
\begin{eqnarray}
\partial^2_{x_1}\sigma=4\pi \delta'(x_1)+{\cal M}^2\sigma\,.
\label{domain wall equation simplified}
\end{eqnarray}
The solution to this equation is given by
\begin{eqnarray}
\sigma(x_1)=2\pi \mbox{sign}(x_1)e^{-{\cal M}|x_1|}\,,
\end{eqnarray}
where $ \mbox{sign}(x_1)$ is the sign function. Substituting $\sigma(x_1)$ back into $\left \langle W(C) \right \rangle$, we find
\begin{eqnarray}
\sigma_{\mbox{\scriptsize mean-field}}=\frac{g_3^2{\cal M}}{4}\,.
\label{sigma mean field appendix}
\end{eqnarray}
This coincides exactly with (\ref{the string tension using the mean field approximation}) obtained using the many-body approach. Comparing (\ref{sigma mean field appendix}) with the exact result (\ref{exact string tension from Polyakov calculations}),  we find that the mean-field calculations, $\sigma_{\mbox{\scriptsize mean-field}}$, are off by a factor of $2/\pi^2$.

\subsection{The perturbative treatment of the Wilson loop}

In this section, we perform a perturbative treatment to the expectation value of the Wilson loop. By doing that, we recover the  second term of (\ref{final effective partition function for the Polyakov loop}), which works as an alternative derivation of the potential between two external probes.  We start from (\ref{ the ultimate integral}), assume that the monopole-instanton density $\rho$ follows a Gaussian distribution, i.e. $\langle \rho(x)\rangle=0$, and expand in  $\rho$ to find
\begin{eqnarray}
\left \langle W(C) \right \rangle=\sum_{n=0}^\infty W^{2n}(C)\,,
\end{eqnarray}
where
\begin{eqnarray}
\nonumber
W^{2n}(C)=\frac{(-1)^n}{(2n)!}\int d^3x_1...d^3x_{2n}\langle \rho(x_1)...\rho(x_{2n}) \rangle \langle \eta(x_1)...\eta(x_{2n}) \rangle\,, 
\label{exact FC in x space}
\end{eqnarray}
and we used the fact that the correlators of an odd number of density operators vanish identically in the Gaussian approximation.
Taking the Fourier transform, we obtain
\begin{eqnarray}
W^{2n}(C)=\frac{(-1)^n}{(2n)!}\frac{1}{(2\pi)^{6n}}\int d^3 p_1...d^3p_{2n}\langle \rho(p_1)...\rho(p_{2n}) \rangle \langle \eta(p_1)\rangle...\langle\eta(p_{2n}) \rangle\,.
\label{exact FC in p space}
\end{eqnarray}
Let us notice that expanding in $\rho$ is a legitimate thing to do since the monopole density is assumed to be low. This is a prerequisite for the validity of the dilute gas approximation that we assume throughout this work.

As we mentioned above, we assume that the density-density correlation function $\langle \rho( x_1)...\rho(x_{2n}) \rangle$ can be obtained within the Gaussian approximation. 
The function $\left \langle \rho(x_1) \rho (x_2) \right\rangle $ can be obtained from (\ref{Polyakov action}) by taking the functional derivative of (\ref{ the ultimate integral}) with respect to $\eta(x_1)$ and $\eta(x_2)$. Then, using the Gaussian approximation (\ref{the Wilson line in the Gaussian approximation}), and taking the functional derivative in the Fourier space, we find
\begin{eqnarray}
\label{polyakov density}
\left \langle \rho(p_1)\rho (p_2) \right\rangle=\left(\frac{g_3}{4\pi}\right)^2\frac{(p^2+p_m^2){\cal M}^2}{p^2+p_m^2+{\cal M}^2}\left(2\pi\right)^3\delta^{(3)}(p_1-p_2)\,.
\end{eqnarray}
Similarly, the $2n$-point density correlator $\langle \rho(p_1)...\rho(p_{2n})\rangle$ can be calculated by taking the $2n$-th functional derivative with respect to some external current source to find
\footnote{We hasten to warn the reader that the density-density correlation function given in \cite{Polyakov:1976fu} is not correct.}
\begin{eqnarray}
\langle \rho(p_1)...\rho(p_{2n})\rangle= \left(\frac{g_3}{4\pi}  \right)^{2n}\sum_{i=1}^{(2n)!/(2^nn!)}\prod_{i \neq j=1}^{n} \left(\frac{{\cal M}^2p_i^2}{p_i^2+{\cal M}^2}\right)(2\pi)^{3n}\delta^{(3)}(p_i-p_j)\,.
\label{density density correlator in p space}
\end{eqnarray}
For example, the four-point function reads
\begin{eqnarray}
\nonumber
\langle \rho(p_1)\rho(p_2)\rho(p_3)\rho(p_4) \rangle&=&(2\pi)^6 \left(\frac{g_3}{4\pi}  \right)^4\left[\left(\frac{{\cal M}^2p_1^2}{p_1^2+{\cal M}^2}\right)\left(\frac{{\cal M}^2p_3^2}{p_3^2+{\cal M}^2}\right)\delta^{(3)}(p_1-p_2)\delta^{(3)}(p_3-p_4)\right.\\
&&\left.+(p_2 \rightarrow p_3, p_3 \rightarrow p_2  ) +(p_2 \rightarrow p_4, p_4 \rightarrow p_2  ) \right]\,.
\end{eqnarray}
Next, we compute $\eta(x)$ as given from (\ref{eta integral}), where the surface $S$ is to be taken in the $x_0-x_2$ plane setting $x_1=0$. In the zero temperature case, we take a loop that extends from $x_2=R_1$ to $x_2=R_2$ and from $x_0=-\infty$ to $x_0=\infty$. Alternatively, we can  perform our calculations at a finite temperature $1/\beta$ and then we take $\beta \rightarrow \infty$.  By doing that, we compactify the $x_0$ direction over a circle of radius $\beta$, and hence we sum over an infinite tower of modes along the $x_0$ direction. In this case, the surface $S$ is a cylinder of radius $1/\beta$ that extends from $R_1$ to $R_2$, see Figure (\ref{cylinder}).  Thus, we have
\begin{eqnarray}
\nonumber
\eta(x)&=&\int_0^\beta \int_{R_1}^{R_2}dy_0dy_2\sum_{n=-\infty}^{\infty} \frac{x_1-y_1}{\left[(x_1-y_1)^2+(x_2-y_2)^2+(x_0-y_0+n\beta)^2 \right]}\left.\right|_{y_1=0}\\
\nonumber
&=&x_1\int_{R_1}^{R_2} dy_2 \int_{-\infty}^{\infty}dy_0\frac{1}{\left[(x_2-y_2)^2+x_1^2+y_0^2\right]^{3/2}}\\
&=&2\tan^{-1}\left[\frac{x_2-R_2}{x_1} \right]-2\tan^{-1}\left[\frac{x_2-R_1}{x_1} \right]\,.
\end{eqnarray}
Using the definition (\ref{the angle theta}), we can write $\eta(x)$ as
\begin{eqnarray}
\eta(x)=2\left(\Theta(\vec R_1-\vec x)-\Theta(\vec R_2-\vec x)\right)\,,
\label{the form of eta in terms of theta}
\end{eqnarray}
where $\vec R_{1,2}=R_{1,2}\hat x_2$, and $\hat x_2$ is a unit vector in the $x_2$-direction. In the limit $R_{1,2} \rightarrow \pm\infty$ we find a $4\pi$ discontinuity in the value of $\eta(x)$ as we cross the $x_1$-axis. Hence, $\eta(x)$, in addition of being the magnetic flux due to a unit magnetic charge, it is also the solid angle seen by a monopole-instanton located at the Euclidean point $x$. 
Now, we take the Fourier transform of $\eta(x)$ to obtain
\begin{eqnarray}
\label{FT of eta}
\eta(\vec p,p_m)&=& \int_0^\beta d\tau \int d^x e^{ip_m\tau}e^{i \vec p\cdot \vec x} \eta(x)
= \left[e^{ip_2 R_2}-e^{ip_2 R_1} \right]\beta \delta_{p_m,0}\tilde A_0^{m(p)}(\vec p,0)\,,
\end{eqnarray}
where
\begin{eqnarray}
\tilde A_0^{m(p)}(\vec p,0)=2\int d^2 x e^{i\vec p\cdot \vec x}\tan^{-1}\left[\frac{x_2}{x_1} \right]=-4\pi\frac{ p_1}{p_2(p_1^2+p_2^2)}\,.
\label{form of A0}
\end{eqnarray}
We note that the choice of the symbol $\tilde A_0^{m(p)}$ to denote the Fourier transform of $\eta (x)$ is not arbitrary. Recalling (\ref{the form of eta in terms of theta}) and (\ref{Fourier transform of A0}), we see that $\tilde A_0^{m(p)}$ is just the Fourier transform of the angle $\Theta (\vec x)$.

Inserting (\ref{FT of eta}) and (\ref{density density correlator in p space}) into (\ref{exact FC in p space}), we find
\begin{eqnarray}
W^{2n}(C)=\frac{(-1)^n}{(2n)!}\frac{(2n)!}{2^nn!}\left[ \int \frac{d^3p}{(2\pi)^3}\frac{1}{L}\left(\frac{g_c}{4\pi}  \right)^2 \left(\frac{{\cal M}^2p^2}{p^2+{\cal M}^2}\right)\eta(p)\eta(-p)  \right]^n\,.
\end{eqnarray}
Since $\eta (p)$ depends only on the two-dimensional vector $\vec p$, we can replace $\int d^3 p/(2\pi)^3$ by $\int d^2p/((2\pi)^2\beta)$, and use (\ref{FT of eta}) to find
\begin{eqnarray}
\nonumber
W(C)&=&\sum_n^\infty\frac{(-1)^n}{(2n)!}\frac{(2n)!}{2^nn!}\left[ \int \frac{d^2p}{(2\pi)^2\beta}\left(\frac{g_3}{4\pi}  \right)^2 \left(\frac{{\cal M}^2p^2}{p^2+{\cal M}^2}\right)\beta \tilde A_0^{m(p)}(\vec p,0)\beta \tilde A_0^{m(p)}(-\vec p,0)\left[2e^{-ip_2(R_2-R_1)}-2 \right]  \right]^n\,\\
\nonumber
&=&\sum_n^\infty\frac{(-1)^n}{n!}\beta^n\left[ \int \frac{d^2p}{(2\pi)^2}\Pi(\vec p,0)\tilde A_0^{m(p)}(\vec p,0)\tilde A_0^{m(p)}(-\vec p,0)\left[e^{-ip_2(R_2-R_1)}-1 \right]  \right]^n\\
&=&\exp\left[-\beta \int \frac{d^2p}{(2\pi)^2}\Pi(\vec p,0)\tilde A_0^{m(p)}(\vec p,0)\tilde A_0^{m(p)}(-\vec p,0)\left[e^{-ip_2(R_2-R_1)}-1 \right] \right]\,,
\end{eqnarray}
where $\Pi(\vec p,0)$ is given by (\ref{the definition of PI}). The quantity inside the exponent exactly matches the second term in (\ref{final effective partition function for the Polyakov loop}).

\section{ Sums and integrals}

In this appendix, we perform the Matsubara sums and integrals in (\ref{main effective action}). First, we consider the function $\psi_0$. Integrating over $\tau$, and taking the derivative with respect to $s$ at $s=0$, we obtain
\begin{eqnarray}
\nonumber
-\frac{d}{ds}\left[\frac{1}{\Gamma(s)} \int _0^\infty d\tau\frac{e^{-\tau(M_W^2+p^2)}}{\sqrt{4\pi \tau}}\tau^{s-1} \psi_0\right]_{s=0}=T\sum_{n}\log\left[p^2+M_W^2+\left(\omega_n+iT\log\Omega\right)^2 \right]\,.\\
\label{the first integral in appendix B}
\end{eqnarray}
Since this sum is divergent, we first differentiate the above expression with respect to $M_W^2$ and perform the sum. Then, we integrate over $p$ to get a function that behaves as $\sim  p$ at $p \rightarrow \infty$. To regularize the integral, we subtract the same form of the integral evaluated at $i\log\Omega=0$. Then, we find
\begin{eqnarray}
\nonumber
&&-\int \frac{d^2p}{\left(2\pi\right)^2}\frac{d}{ds}\left[\frac{1}{\Gamma(s)} \int _0^\infty d\tau\frac{e^{-\tau(M_W^2+p^2)}}{\sqrt{4\pi \tau}}\tau^{s-1} \psi_0\right]_{s=0}\\
\nonumber
&=&-\frac{T}{2\pi}\int^{y=M_W} dy y \log\left[\frac{\cosh\frac{y}{T}-\cos\left(i\log \Omega\right)}{\cosh\frac{y}{T}-1} \right]\\
&\underbrace{\cong}_{T<<M_W}&\frac{T^2 M_W}{\pi}e^{-M_W/T}\left(1-\cos\left(i\log\Omega\right)\right)\,.
\label{psi0}
\end{eqnarray}  

Next, we turn to the term $\tau^2\psi_0$. Performing the integral over $\tau$, taking the derivative with respect to $s$ at $s=0$, performing the integral over $p$, and then summing over the Matsubara frequencies, we find
\begin{eqnarray}
\nonumber
&&-\int \frac{d^2p}{\left(2\pi\right)^2}\frac{d}{ds}\left[\frac{1}{\Gamma(s)} \int _0^\infty d\tau\frac{e^{-\tau(M_W^2+p^2)}}{\sqrt{4\pi \tau}}\tau^{s-1} \tau^2\psi_0\right]_{s=0}\\
\nonumber
&&=-\frac{\sinh\frac{M_W}{T}}{8\pi M_W\left(\cosh\frac{M_W}{T}-\cos\left(i\log \Omega\right)\right)}\\
&&\underbrace{\cong}_{T<<M_W}-\frac{1}{8\pi M_W}\left[1+2e^{-M_W/T}\cos\left(i\log\Omega\right) \right]\,.
\label{tau2psi0}
\end{eqnarray}

Finally, by repeating the same steps we followed to calculate $\tau^2\psi_0$, we obtain for $\tau^2\psi_2$
\begin{eqnarray}
\nonumber
&&-\int \frac{d^2p}{\left(2\pi\right)^2}\frac{d}{ds}\left[\frac{1}{\Gamma(s)} \int _0^\infty d\tau\frac{e^{-\tau(M_W^2+p^2)}}{\sqrt{4\pi \tau}}\tau^{s-1} \tau^2\psi_2\right]_{s=0}\\
\nonumber
&&=\frac{1}{16\pi M_W}\left\{\frac{2\sinh\frac{M_W}{T}}{\cosh\frac{M_W}{T}-\cos\left(i\log \Omega\right)}+\frac{\cos\left(i\log \Omega\right)\left(\sinh\frac{M_W}{T}-\frac{M_W}{T}\cosh\frac{M_W}{T}\right)-\frac{1}{2}\sinh\frac{2M_W}{T}+\frac{M_W}{T}}{\left(\cosh\frac{M_W}{T}-\cos\left(i\log \Omega\right)\right)^2} \right\}\\
&&\underbrace{\cong}_{T<<M_W}\frac{1}{16\pi M_W}\left[1-2\frac{M_W}{T}e^{-M_W/T}\cos\left(i\log\Omega\right) \right]\,.
\label{tau2psi2}
\end{eqnarray}
%

\section{Performing the path integral using the duality transformation prescription}

In this appendix, we work out the details that lead us to the partition function (\ref{final expression for Z}) starting from the auxiliary action 
\begin{eqnarray}
\nonumber
S_{\mbox{\scriptsize aux}}&=&\int d^{2+1}x\frac{1}{4g_{3\mbox{\scriptsize eff}}^2}\left({\cal F}_{\mu\nu}\right)^2+\frac{1}{2g_{3\mbox{\scriptsize eff}}^2}{\cal F_{\mu\nu}}F^{\mbox{\scriptsize ph}}_{\mu\nu}+ \frac{g_{3\mbox{\scriptsize eff}}^2}{2}\Phi_\mu^2\\
&&+i\epsilon_{\mu\nu\lambda}\partial_\mu A^{\mbox{\scriptsize ph}}_\nu \Phi_\lambda-i\sum_A q_AA_0^{\mbox{\scriptsize ph}}(\vec x,x_0)\delta^{(2)}(\vec x-\vec R_A)\,.
\label{auxiliary Lagrangian in appendix}
\end{eqnarray}
Varying (\ref{auxiliary Lagrangian in appendix})  with respect to $A_\mu^{\mbox{\scriptsize ph}}$, we obtain
\begin{eqnarray}
\frac{1}{g_{3\mbox{\scriptsize eff}}^2}\partial_\alpha {\cal F}_{\alpha \beta}+i\epsilon_{\alpha \beta \gamma }\partial_\alpha\Phi_\gamma=-i\sum_{A}q_A\delta^{(2)}\left(\vec x-\vec x_A\right)\delta_{0\beta}\,.
\label{equation of motion for Phi}
\end{eqnarray}
The solution to the equation of motion (\ref{equation of motion for Phi}) is given by
\begin{eqnarray}
\Phi_\mu=\frac{i}{g_{3\mbox{\scriptsize eff}}^2}{\cal B}_\mu +b_\mu\,,
\label{solution to Phi}
\end{eqnarray}
where ${\cal B}_\mu=\epsilon_{\mu\nu\alpha}{\cal F}_{\nu\alpha}/2$. The vector field $b_\mu$ can be decomposed into curl-free and divergence-free parts
\begin{eqnarray}
b_\lambda =\partial_\lambda \sigma+\epsilon_{\lambda \mu \nu}\partial_{\mu}C_{\nu}\,,
\label{decomposition of b}
\end{eqnarray}
where $\partial_\nu C_\nu=0$. Substituting (\ref{solution to Phi}) and (\ref{decomposition of b}) into (\ref{equation of motion for Phi}), we find that the $\sigma$ parts drop out, while the equation of $C_\mu$ reads
\begin{eqnarray}
\nabla^2 C_\mu=-\sum_A q_A \delta_{0\mu}\delta^{(2)}\left(\vec x-\vec x_A  \right)\,,
\end{eqnarray}
where the Laplacian $\nabla^2$ is defined on $\R^2\times \S^1_\beta$. At this stage, let us define the Green's function $G(\vec x-\vec x',x_0-x_0')$, which satisfies
\begin{eqnarray}
\nabla^2 G(\vec x-\vec x',x_0-x_0')=-\delta^{(2)}(\vec x-\vec x_a)\delta(x_0-x_{0a})\,.
\label{greens function equation}
\end{eqnarray}
The solution to (\ref{greens function equation}) on $\R^2\times \S^1_\beta$ is
\begin{eqnarray}
G(\vec x-\vec x',x_0-x_0')=\frac{1}{4\pi}\sum_{n=-\infty}^{\infty}\frac{1}{\sqrt{\left(\vec x-\vec x'\right)^2+(x_0-x_0'+n\beta)^2}}\,.
\end{eqnarray}
Then, the solution $C_\mu$ is given by
\begin{eqnarray}
\nonumber
C_\mu&=&\int d^{2+1} x ' \sum_A q_A \delta_{0\mu}\delta^2 \left(\vec x'-\vec x_A\right) G(\vec x-\vec x',x_0-x_0')\\
\nonumber
&=& \frac{1}{4\pi}\sum_A q_A\delta_{0\mu}\sum_{n=-\infty}^{\infty}\int_0^\beta dx_0 \frac{1}{\sqrt{\left(\vec x-\vec x'\right)^2+(x_0-x_0'+n\beta)^2}}\\
&=&-\frac{\delta_{0\mu}}{2\pi}\sum_A q_A \log \left|\vec x-\vec x_A\right|\,.
\end{eqnarray}
Now, the solution $\Phi_\mu$ reads
\begin{eqnarray}
\Phi_\mu=\frac{i}{g_{3\mbox{\scriptsize eff}}^2}{\cal B}_\mu+\partial_\mu\sigma +K_\mu\,,
\label{final expression of Phi}
\end{eqnarray}
where
\begin{eqnarray}
K_\mu=-\frac{1}{2\pi}\epsilon_{\mu\nu0}\sum_A q_A \partial_\nu\log \left|\vec x-\vec x_A\right|\,.
\end{eqnarray}
Next, we substitute $\Phi_\mu$ into (\ref{auxiliary Lagrangian in appendix}) and use integration by parts to obtain
\begin{eqnarray}
S_{\mbox{\scriptsize aux}}=\int d^{2+1} x \frac{g_{3\mbox{\scriptsize eff}}^2}{2}\left(\partial_\mu\sigma\right)^2+\frac{g_{3\mbox{\scriptsize eff}}^2}{2}K_\mu^2-i\sigma \partial_\nu {\cal B}_\nu+i{\cal B}_\mu K_\mu-g_{3\mbox{\scriptsize eff}}^2\sigma \partial_\mu K_\mu\,.
\label{the simplified aux action}
\end{eqnarray}
The term $ \partial_\mu K_\mu$ is obviously zero due to the anti-symmetry of $\epsilon_{0\mu\nu}$, while $\partial_\mu {\cal B}_\mu=4\pi\sum_{a}q_a\delta^{(3)}(x-x_a)$. Next, we turn to the calculations of the other terms (\ref{the simplified aux action}).  The monopole-instanton field ${\cal B}_\mu$ is
\begin{eqnarray}
{\cal B}_\mu=\sum_a q_a \left(\frac{(x-x_a)_\mu}{\left|x-x_a \right|^3}\right)^{(p)}\,,
\end{eqnarray}
where the superscript $(p)$, as usual, denotes the periodicity along $\S^1_\beta$, which can be enforced by summing an infinite number of images along the circle. Then we have 
\begin{eqnarray}
\nonumber
\int d^{2+1} x {\cal B}_\mu K_\mu&=&-\int  d^{2+1} x \sum_{aA} q_aq_A \int_0^\beta \sum_{n=-\infty}^{\infty}\int d^2 x \frac{\epsilon_{ij}(x-x_a)_i(x-x_A)_j}{\left[(\vec x-\vec x_a)^2+(x_0-x_{0a}+n\beta)^2 \right]^{3/2}|\vec x-\vec x_A|^2}\\
&=&-2\sum_{aA}q_aq_A\int d^2 x  \frac{\epsilon_{ij}(x-x_a)_i(x-x_A)_j}{|\vec x-\vec x_a|^2|\vec x-\vec x_A|^2}\,,
\end{eqnarray}
which is zero under symmetric integration. Then, the term $K_\mu^2$ is
\begin{eqnarray}
\nonumber
\int d^{2+1} x K_\mu^2&=& \frac{1}{\left(2\pi\right)^2T}\sum_{AB}q_Aq_B \int d^2 x \partial_i \log \left|\vec x-\vec x_A\right|\partial_i \log \left|\vec x-\vec x_A\right|\\
\nonumber
&=&\frac{1}{4\pi^2 T}\sum_{AB}q_Aq_B \int d^2 x \frac{(x-x_a)_i(x-x_A)_i}{|\vec x-\vec x_a|^2|\vec x-\vec x_A|^2}\\
&=&-\frac{1}{2\pi T}\sum_{AB}q_Aq_B \log T|\vec x_A-\vec x_B|\,,
\end{eqnarray}
which is the Coulomb potential between W-bosons. Finally,  varying (\ref{the simplified aux action}) with respect to $\sigma$, we obtain
\begin{eqnarray}
g_{3\mbox{\scriptsize eff}}^2 \nabla^2\sigma=-i4\pi\sum_a q_a\delta^{(3)}(x-x_a)\,,
\end{eqnarray}
which is solved by
\begin{eqnarray}
\sigma= \frac{i4\pi}{g_{3\mbox{\scriptsize eff}}^2}\sum_a q_a G(\vec x-\vec x_a, x_0-x_{0a})\,.
\end{eqnarray}
Collecting everything and substituting in (\ref{the simplified aux action}), we find
\begin{eqnarray}
S_{\mbox{aux}}=\frac{8\pi}{g_{3\mbox{\scriptsize eff}}^2}\sum_{ab}q_aq_b G(\vec x_a-\vec x_b, x_{0a}-x_{0b})-\frac{g_{3\mbox{\scriptsize eff}}^2}{4\pi T}\sum_{AB}q_Aq_B \log T|\vec x_A-\vec x_B|\,.
\end{eqnarray} 
Thus, the full partition function reads
\begin{eqnarray}
\nonumber
{\cal Z}_{\mbox{\scriptsize grand}}&=&\sum_{N_{m\pm}, q_a=\pm 1 }\sum_{N_{W\pm}, q_A=\pm 1 }\frac{\xi_m^{N_{m+} + N_{m-}}}{N_{m+}! N_{m-}!}\frac{(T\xi_W)^{N_{W+} + N_{W-}}}{N_{W+}! N_{W-}!}\prod_a^{N_{m+} + N_{m-}} \int d^{2+1}x_a\prod_A^{N_{W+} + N_{W-}} \int \frac{d^{2}x_A}{T}\\
\nonumber
&&\times \exp \left[-\frac{8\pi^2}{g_{3\mbox{\scriptsize eff}}^2}\sum_{a,b}q_aq_bG(x_a- x_b)+\frac{g_{3\mbox{\scriptsize eff}}^2}{4\pi T}\sum_{A,B}q_Aq_B\log T|\vec x_A-\vec x_B|+2i\sum_{aA}q_aq_A\Theta\left(\vec x_a-\vec x_A\right)\right]\,,\\
\label{final expression for Z in the appendix}
\end{eqnarray}
which is valid in the range $0 \leq T<M_W$.

\end{document}